\documentclass[pra,aps,twocolumn,show
pacs,nofootinbib,floatfix]{revtex4-1}
\usepackage{amsmath}
\usepackage{amssymb}
\usepackage{graphicx}                       
\usepackage{dcolumn}                        
\usepackage{bm}                             
\usepackage{color}
\usepackage{epstopdf}

\begin{document}

\newcommand{\be}{\begin{equation}}
\newcommand{\ee}{\end{equation}}
\newcommand{\bea}{\begin{eqnarray}}
\newcommand{\eea}{\end{eqnarray}}
\newcommand{\me}{m_{\rm e}}                 
\newcommand{\mn}{m_{\rm n}}                 
\newcommand{\HCM}{H_{\rm CM}}
\newcommand{\Hcoupl}{H_c}
\newcommand{\Hrel}{H_{\rm rel}}
\newcommand{\D}{\mathcal D}                 
\newcommand{\Lt}{\mathcal L}                
\newcommand{\Hf}{{\mathcal H}}              
\newcommand{\E}{\varepsilon}                
\newcommand{\N}{{\mathcal N}}
\newcommand{\vg}[1]{\mbox{\boldmath $#1$}}
\newcommand{\vv}[1]{\mathbf #1}
\newcommand{\es}{\varepsilon_s}             
\newcommand{\Meff}{M_{\rm eff}}             

\title{Cyclotron transitions of bound ions} 

\author{Victor~G.~Bezchastnov$^{1,2}$ and George G. Pavlov$^3$}

\affiliation{$^1$Ioffe Physical-Technical Institute, 194021 St.-Petersburg, Russia}
\affiliation{$^2$Institute of Physical Chemistry, University of Heidelberg, INF 229, 
                 69120 Heidelberg, Germany}
\affiliation{$^3$Pennsylvania State University, 525 Davey Lab,
                 University Park, PA 16802, U.S.A.}

\date{\today}

\begin{abstract}

A charged particle in a magnetic field possesses discrete energy levels associated with 
particle's rotation around the field lines. The radiative transitions between these 
levels are the well-known cyclotron transitions. We show that a bound complex of 
particles with a non-zero net charge displays analogous transitions 
between the states of confined motion of the entire complex in the field. 
The latter {\em bound-ion cyclotron transitions} are affected by a coupling between 
the collective and internal motions of the complex and, as a result, differ 
from the transitions of a ``reference'' bare ion with the same mass and charge. 
We analyze the cyclotron transitions for complex ions by including the coupling within 
a rigorous quantum approach. Particular attention is paid to comparison of
the transition energies and oscillator strengths to those of the bare ion. 
Selection rules based on integrals of collective motion are derived for 
the bound-ion cyclotron transitions analytically, and the perturbation and 
coupled-channel approaches are developed to study the transitions quantitatively. 
Representative examples are considered and discussed for positive and negative atomic 
and cluster ions. 

\end{abstract}


\maketitle

\section{Introduction}

Cyclotron radiation is a well-known phenomenon associated with the motion of a charged 
particle in a magnetic field. The particle rotates around the field lines and can 
thereby emit or absorb radiation of the frequency of rotation~\cite{LanLif_1981}. The 
corresponding energy is the {\em cyclotron energy} given for a non-relativistic 
particle by the relation 
\be 
\Omega = |Q|B/M,
\label{Omega}
\ee 
where $Q$ and $M$ are the particle's charge and mass, respectively, and $B$ is the 
strength of the magnetic field. We assume the field to be uniform and adopt the 
atomic system of units in which the unit of magnetic field strength is 
$2.3554 \times 10^5$~T. 

For a stuctureless (bare) ion, such as an atomic nucleus, the properties of the 
cyclotron transitions are fully determined by the ion's mass and charge. 
A complex ion (i.e., a bound system of particles with a nonzero net charge) can 
also rotate as a whole around the field lines and absorb/emit cyclotron photons. 
We will refer to the corresponding radiative transitions as the 
{\em bound-ion cyclotron transitions}. In weak magnetic fields they can be 
approximately considered as cyclotron transitions of a {\em reference} bare ion 
with the mass and charge equal to those of the entire system. Such a description 
is applicable when the reference cyclotron energy $\Omega$ is much lower than 
the ion binding energy, 
which is often the case in various laboratory experiments or in the terrestrial and 
solar magnetospheres. However, in very strong magnetic fields or for loosely 
bound ions, when $\Omega$ is not a negligible fraction of the 
binding energy, the internal structure does affect the bound-ion cyclotron 
transitions. We will focus on such effects in this paper.

A rigorous theoretical description of the bound-ion cyclotron transitions has to 
deal with the coupling of the collective motion to the internal degrees of 
freedom for many-particle systems. In the presence of a magnetic field, this 
coupling is known to be rather nontrivial (e.g. Refs.~\cite{AHS_1981,Johnson1983}) 
but it is of fundamental importance for understanding the properties of atoms and 
molecules in  external fields. 

Generally, the electronic structure and properties of atomic and molecular 
systems in external magnetic fields have been intensively studied for already a 
few decades, see, e.g., Ref.~\cite{Ruder_1994} and references 
therein. Many of these studies have been performed assuming the nuclei to be 
infinitely heavy, i.e. neglecting the collective motion and considering the 
electronic configuration only. The coupling between the internal and 
center-of-mass (c.m.) motions in magnetic fields has been examined in 
Refs.~\cite{Johnson1983,Schmelcher1991} and shown to 
influence the quantum structure for many-body systems with both zero and 
non-zero net charge (see, e.g., Refs.~\cite{Schmelcher1994,Schmelcher1997}). 
For neutral systems, the c.m.\ motion can be separated out by exploiting a 
specific many-body integral of motion in a magnetic field, the total 
pseudomomentum~\cite{Johnson1983}. This procedure, called the pseudo-separation 
because the resulting Hamiltonian for the internal motions depends on the conserved 
pseudomomentum, facilitates accounting for the coupling in the quantum structure 
calculations. For charged systems, the nature of the coupling does not allow a 
separation of the c.m.\ motion~\cite{Johnson1983,Schmelcher1991} and makes the 
quantum structure calculations particularly tedious. To study the interaction 
between the c.m.\ and the internal motions, the classical dynamics approaches 
have been applied in a series of 
studies~\cite{Schmelcher1997,Schmelcher_1992_1995,Leitner1998,Melezhik1999,Schmelcher2000,BSC2002}. 

Quantum calculations with account for the coupling between the collective and 
internal motions in magnetic fields require a substantial theoretical effort 
already for two-body charged systems. By now, the quantum states for ions moving 
in a magnetic field have been comprehensively investigated for hydrogen-like ions, 
in particular He$^+$~\cite{BaVi86, BezPavVen98} and for ions that can be considered 
as one-electron systems. An example of such a system is, e.g., the 
magnetically-induced negative ions in which the excess electron is bound exclusively 
due to the presence of the magnetic field~\cite{BSC2000,BSC-review,BSC2007}. The 
states were rigorously computed as discrete eigenstates of the integrals of 
collective motion of the ions. For hydrogen-like ions, the impact of the motion 
on quantum states becomes prominent for field strengths above $10^8-10^9$~T typical 
for the neutron star environments. The motion of the magnetically-induced atomic 
anions affects the bound states at field strengths of the order of tens of Tesla, 
accessible in terrestrial laboratories. 

As far as the radiative transitions are concerned, the majority of the 
performed studies conventionally focus on the transitions between the internal 
states and do not address the transitions between the states of collective 
motion. Many results on the internal electronic transitions 
were obtained by neglecting the coupling to c.m., in particular for the hydrogen 
atom~\cite{Friedrich1983,Greene1983,Meinhardt1999,Vieyra2008} and for the atomic 
helium~\cite{Al-Hujaj2003,Medin2008}. With account for the coupling, the 
electronic transitions have been studied for the hydrogen atom - see, e.g., 
Refs.~\cite{Wunner1981,Cuvelliez1992,Pavlov1995} on the bound-bound transitions 
and Refs.~\cite{BP1994,Potekhin1997} on the bound-free transitions. 
For ions the effect of coupling on radiative transitions has not yet been 
analyzed so systematically. Some results on the bound-ion cyclotron transitions 
have been presented in Ref.~\cite{PB2005} for the He$^+$ ions and in 
Ref.~\cite{BSC2007} for the magnetically-induced atomic anions. 

The bound-ion cyclotron transitions are primarily contributed by the 
transitions between the states of collective motion. We will therefore start 
from a general analysis of transitions involving the c.m.\ and internal degrees 
of freedom coupled in a magnetic field. We proceed then to a perturbation 
treatment of the coupling followed by numerical studies of the bound-ion 
cyclotron transitions. 

The paper is organized as follows. A basic quantum description of motion 
for a single charge and a system of charges in a magnetic field is outlined, 
and general selection rules for the dipole radiative transitions are provided 
in Section~II. In Section~III we consider an atomic ion in a 
magnetic field. A perturbation approach to the coupling between the c.m.\ 
and internal motions is employed to identify the selection rules for the 
bound-ion cyclotron transitions and to quantify the difference of the radiation 
energies and oscillator strengths from those of bare ions. 
An important quantity obtained in this section is the effective mass dependent 
on the internal quantum states of the ion. 
Section~IV presents a coupled-channel approach to the bound-ion 
cyclotron transitions. The corresponding numerical results are described 
in Section~V for positive helium ions in strong astrophysical magnetic 
fields and for the negative magnetically-induced ions formed by the xenon 
and argon atoms and clusters in a magnetic field which can be maintained in 
laboratories. The properties of the transitions 
are discussed and compared with the results of the perturbation treatment. 
Concluding remarks are given in Section~VI. The relevant mathematical details 
are provided in Appendix~A on the quantum description of the cyclotron 
transitions of a bare ion, in Appendix~B on the integrals of motion and 
dipole selection rules, in Appendix~C on the perturbation analysis and in 
Appendix~D on the coupled-channel calculations of the bound-ion states and 
cyclotron transitions.

\section{Basic relations for ions in a magnetic field}

Quantum description of the motion in magnetic fields for a single charge and for a 
system of charges can be found in many textbooks and original papers (see, e.g., 
Refs.~\cite{LanLif_1981,Johnson1983,AHS_1981,Ruder_1994}). To study the cyclotron 
transitions for complex ions, we apply the non-relativistic quantum approach 
which does not include the particle spins as they do not affect 
the transition energies and oscillator strengths\footnote{By excluding spins we 
do not account for, e.g., magnetic nuclear resonance transitions, which are 
magnetic dipole transitions much weaker than the electric dipole transitions 
considered in this paper.}. We choose the z-axis of the 
coordinate frame along the magnetic field and employ the symmetric gauge 
for the vector-potential, ${\bf A} = (1/2)\,{\bf B}\times{\bf r}$.  

\subsection{Bare ions}

It is instructive to compare the properties of the cyclotron transitions for 
complex ions with those of a bare structureless ion. The motion of the 
bare ion with the mass $M$ 
and charge $Q$ in the plane ${\bf R}_\perp = (X,Y,0)$ perpendicular to the magnetic 
field is described by the Hamiltonian 
\be
H_1 = \frac{1}{2M} \left( {\bf P}_\perp 
                        - \frac{Q}{2}\,{\bf B}\times{\bf R}_\perp \right)^2,
\label{H1}
\ee
where ${\bf P}_\perp$ is the canonical momentum. The eigenvalues of 
the Hamiltonian~(\ref{H1}) are 
the discrete {\em Landau energies}
\be
E^{\rm Lan}_N(\Omega) = (\Omega/2)(2N+1)~,
\label{Lan-en}
\ee
where $N = 0,1,2,\ldots$ enumerates the equidistant Landau levels separated by 
the cyclotron energy $\Omega$. 
Rotation of the ion around the magnetic filed lines exhibits two {\em commuting} 
integrals of motion: the square of the transverse pseudo-momentum, $K_\perp^2$, 
and the longitudinal component of the angular momentum, $L_z$; see Eqs.~(\ref{K}) 
and (\ref{L}). The respective discrete eigenvalues are determined by the integers 
$N_0$ and $L$ (see Eqs.~(\ref{K-vals}) and (\ref{L-vals})), with the sums
\be
N_0 + L = N 
\label{relation-numbers}
\ee
equal to the Landau level numbers. 

The radiative transitions between the Landau levels are the ion 
{\em cyclotron transitions} determined by the dipole operators 
\be
D^{(\beta)} = Q(X+{\rm i}\,\beta\,Y)/\sqrt{2}~,
\label{dipop}
\ee
where $\beta = +1$ and $\beta = -1$ correspond to the right and left circular 
polarizations, respectively. The dipole matrix elements differ from zero for 
the transitions $N \to N'$ between the neighboring levels, see Eq.~(\ref{D-cyc}). 
The corresponding {\em selection rules}, {\em transition energies} and 
{\em oscillator strengths} are 
\bea
N' &=& N - \beta\sigma~,
\label{selection-cyc}
\\
\omega^{\rm cyc}_{N',N} &=& -\beta\sigma\,\Omega~,
\label{omega-cyc}
\\
f^{\rm cyc}_{N',N} &=& -\beta\sigma\,\frac{Q^2}{M}\,\left( N' + N + 1 \right)~,
\label{f-cyc}
\eea
where $\sigma = \pm 1$ is the sign of the charge $Q$. 

\subsection{System of particles}

We turn now to a system of particles in a magnetic field. 
The Hamiltonian for the system is 
\be
H = \sum_a 
    \frac{1}{2m_a} \left( {\bf p}_a - \frac{e_a}{2}\,{\bf B}\times{\bf r}_a \right)^2 
  + V~,
\label{H-many}
\ee
where {\em a} labels the particles with the masses $m_a$, charges $e_a$, 
coordinates ${\bf r}_a$ and momenta ${\bf p}_a$, and the potential $V$ is the sum of 
pairwise Coulomb potentials. Already for the system of two particles, calculations 
of the quantum states of the Hamiltonian~(\ref{H-many}) require a 
numerical treatment. Still, the quantum states can be determined as the eigenstates 
of two commuting 
integrals of motion, $K_\perp^2$ and $\Lt_z$, that are now the square of the 
transverse component and the longitudinal component of the {\em total} pseudo- and 
angular momenta, respectively. These quantities specify the {\em collective} motion 
for the system, see Appendix~B for details. The discrete eigenvalues 
$K_\perp^2$ and $\Lt_z$ are determined by the integers $N_0=0,1,2,\ldots$ and 
${\cal L}=0,\pm 1,\pm 2,\ldots$, respectively 
(see Eqs.~(\ref{K-vals}) and (\ref{L-tot-vals})). 
As proved below, the quantum energies are determined by the sum of these two numbers, 
\be
{\cal N} = N_0 + \Lt~,
\label{relation-numbers-1}
\ee
and are degenerate with respect to $N_0$. 

Notice that the many-particle pseudo-momentum can be transformed to a one-particle 
form (see Appendix~B). Therefore, we use the notations, $K_\perp^2$ and $N_0$, 
for the integral of motion and its quantum number, same as for a bare ion. 
In contrast, the angular momentum of a system of ions is essentially a 
many-particle operator, 
and we introduce the notations $\Lt_z$ and $\Lt$ different from $l_z$ and 
$s$ used for the bare ion. 

In contrast to the bare ion, a many-particle system in the magnetic field 
exhibits the radiative transitions of a broader variety than the ``pure'' cyclotron 
ones. The transitions between the internal states and between the 
states of the collective motion influence each other because the two types of 
motion are coupled in the magnetic field. 

To identify the bound-ion cyclotron transitions, we consider the transitions 
determined by the cyclic components 
\be
\D^{(\beta)}=(\D_x+{\rm i}\,\beta\,\D_y)/\sqrt{2}
\label{dip}
\ee
of the dipole moment of the system, 
\be 
\bm{\D} = \sum_a e_a {\bf r}_a~. 
\label{DipMom} 
\ee 
The corresponding selection rules with respect to the quantum numbers $N_0$ and 
$\cal L$ are $N'_0=N_0$ and ${\cal L}' = {\cal L} - \beta\sigma$. They are 
derived analytically (see Appendix~B) and are general in the sense that they are 
related to the collective motion of the system. These two rules yield 
\be
{\cal N}' = {\cal N} - \beta\sigma~, 
\label{selection-cyc-a}
\ee
in complete analogy with the selection rules~(\ref{selection-cyc}) for the 
cyclotron transitions of the reference bare ion. 
Additional selection rules are related to changes in the internal structure of the 
ion. We study these rules, along with the transition energies and oscillator 
strengths, in the following parts of the paper. 

\section{Perturbation approach}

In this section we consider an atomic ion, for which the analysis of the bound-ion 
cyclotron transitions is facilitated by conservation of the longitudinal angular 
momentum for the isolated electronic configuration. The Hamiltonian for the ion, 
convenient for analysis of the interaction between the c.m.\ and internal motions 
in a magnetic field, has been derived in Ref.~\cite{Schmelcher2000}. A series of 
the canonical and gauge transformations allows one to present the Hamiltonian as 
the sum of three terms that refer to the c.m.\ motion, the internal motion, and 
the coupling of these motions,  
\be
H = H_1 + H_2 + H_3~.
\label{H}
\ee
The Hamiltonian $H_1$ is given by Eq.~(\ref{H1}), where $\vv{R}_\perp$ and 
$\vv{P}_\perp$ are now the coordinate and momentum for the c.m.\ motion. 
Thus, this Hamiltonian describes the reference structureless 
ion with the Landau energies~(\ref{Lan-en}).

The Hamiltonian $H_2$ involves only the coordinates and momenta for the 
motion of the electrons relative to the nucleus. The explicit (rather cumbersome) 
expression for $H_2$ can be found in Ref.~\cite{Schmelcher2000}. 
The property of this Hamiltonian important for our analysis is 
the {\em axial symmetry} with respect to the direction of the magnetic field. 
Similar to the infinitely heavy atomic ion, the electronic states 
described by the term $H_2$ possess the integral of motion given by the 
longitudinal component $l_z$ of the electronic angular momentum, 
\be
(0,0,l_z) = \sum_i \vv{r}_{\perp{i}}\times\vv{p}_{\perp{i}}~, 
\label{l_z} 
\ee
where $i$ enumerates the electrons with the transverse coordinates 
$\vv{r}_{\perp{i}}$ and momenta $\vv{p}_{\perp{i}}$. The eigenstates of 
$H_2$ can therefore be attributed to the discrete values 
\begin{align}
l_z& = -s~,& s& = 0,\pm 1,\pm 2,\ldots~.
\label{l-vals}
\end{align}
We denote the corresponding electronic energies by $\E_{s\nu}$, where $\nu$ 
stands for all quantum numbers other than $s$. 
Note that the $\E_{s\nu}$ values are generally different from the energies for 
the infinitely heavy ion because the Hamiltonian $H_2$ partially accounts for 
the finite ion mass (in particular, by the mass-polarization terms). 
However, the difference is small due to the small ratio of the electron 
and nucleus masses. 
 
The coupling term 
\be
H_3 = \frac{\alpha}{M}\,
      \left[ {\bf B} \times 
             \left( {\bf P}_\perp - \frac{Q}{2}\,{\bf B}\times{\bf R}_\perp \right)
      \right]\,\sum_i {\bf r}_{\perp{i}}~,
\label{H3}
\ee
where $\alpha = 1 + (Q/M)$, has a form of the Stark-type interaction of the 
electrons with an electric field induced by the c.m.\ motion. 

The transformations employed to derive the Hamiltonian~(\ref{H}) yield the 
dipole operators~(\ref{dip}) as the sums of two terms
\be
\D^{(\beta)} = D^{(\beta)} + \alpha\, d^{(\beta)}~,
\label{dipsplit}
\ee
where $D^{(\beta)}$ is the dipole moment~(\ref{dipop}) for the reference ion, and 
\be
d^{(\beta)} = -\sum_i(d_{xi}+{\rm i}\,\beta\,d_{yi})/\sqrt{2}
\label{dip-el}
\ee
is the dipole moment for the electronic subsystem.

To study the impact of the coupling $H_3$ on the bound-ion cyclotron transitions, 
we consider below the cases where the coupling is neglected and where it is treated 
as a perturbation in the Hamiltonian~(\ref{H}). 

\subsection{Zero-order approximation}

The c.m.\ and internal electronic motions are 
independent if the coupling $H_3$ is neglected. 
The zero-order ion energies are the sums of the eigenenergies for the 
c.m.\ term $H_1$ and the electronic term $H_2$, 
\be
E_{\N{s}\nu} = E^{\rm Lan}_N(\Omega) + \E_{s\nu}~.
\label{ens-zero}
\ee
The zero-order wave functions are given by Eq.~(\ref{wfs-zero}) as the products 
of the c.m.\ Landau functions $|N_0,N\rangle$ and the electronic orbitals 
$|s,\nu\rangle$. 

Since the zero-order Hamiltonian commutes with the integrals of motion $K_\perp^2$ 
and $\Lt_z$, the zero-order states can be specified by the quantum 
numbers $N_0$ and $\Lt$. The total pseudo-momentum, transformed in the course of 
the derivations of the Hamiltonian~(\ref{H}), acquires the {\em one-particle} 
form~(\ref{K}) solely involving the c.m.\ coordinate and momentum. Thus, the 
c.m.\ Landau functions in Eq.~(\ref{wfs-zero}) specified by the numbers $N_0$ 
ensure that the zero-order functions are the eigenfunctions of $K_\perp^2$. 
The transformed total angular momentum of the ion is the sum of the c.m.\ and 
electronic parts. The longitudinal component is 
\be
\Lt_z = L_z + l_z~,
\label{Lz-ion}
\ee 
where $L_z$ is determined by the c.m.\ degrees of freedom according to 
Eq.~(\ref{L}), and $l_z$ is given by Eq.~(\ref{l_z}). 
Since the electronic orbitals are attributed to the eigenvalues~(\ref{l-vals}), 
the relation 
\be
N = \N - \sigma{s}
\label{N-ion}
\ee 
assigns the quantum number $\Lt$ to the zero-order states. 
As a result, the ion energies are determined by the number $\N = N_0 + \Lt$, 
being degenerate with respect to $N_0$. 

The radiative transitions between the zero-order states split into the independent 
c.m.\ and electronic transitions determined by the dipole operators $D^{(\beta)}$ 
and $d^{(\beta)}$, respectively. The c.m.\ transitions are the pure 
cyclotron transitions of the reference ion. The corresponding selection 
rules, $N' = N - \beta\sigma$, are the same as for the single ion 
(see Eq.~(\ref{selection-cyc})). 
The selection rules for the electronic transitions can be derived by considering 
the dipole matrix elements $\langle s'\nu' | d^{(\beta)} | s\nu \rangle$. 
From the commutator relation $[l_z,d^{(\beta)}]=\beta d^{(\beta)}$ it follows that 
\be
s' = s + \beta~.
\label{s-selection}
\ee

With account for Eq.~(\ref{N-ion}), the general selection rule 
$\N'=\N-\beta\sigma$ is satisfied by 
two pairs of conditions for the numbers $N$ and $s$. One pair,
\begin{align}
N'& = N - \beta\sigma& s'& = s~,
\label{pair_1}
\end{align}
corresponds to the c.m.\ cyclotron transitions preserving the longitudinal 
angular momentum for the electronic configuration. The other pair,
\begin{align}
N'& = N&               s'& = s + \beta~.
\label{pair_2}
\end{align}
is related to the electronic transitions preserving the state of the 
c.m.\ motion. 
If the first type of the transitions additionally preserves the numbers 
$\nu$ of the electronic states, the transition energies are equal to 
the differences in the c.m.\ cyclotron energies. Thus, the selection rules 
\begin{align}
\N'& = \N - \beta\sigma& s'& = s& \nu'& = \nu
\label{bi-cyc}
\end{align}
identify the cyclotron transitions for the bound ion, and the number 
$N=\N-\sigma{s}$ plays a role of the Landau level number for this ion. 
The transition energies $\omega_{\N'\!,\,\N}$ and oscillator 
strengths $f_{\N'\!,\,\N}$, obtained from the ion energies~(\ref{ens-zero}) 
and wave functions~(\ref{wfs-zero}), are the same as for the reference ion:
\begin{align}
\omega_{\N'\!,\,\N}& = \omega^{\rm cyc}_{N'\!,N}~,&
f_{\N'\!,\,\N}& = f^{\rm cyc}_{N'\!,N}~,
\label{cyc-zero}
\end{align}
where $\omega^{\rm cyc}_{N'\!,N}$ and $f^{\rm cyc}_{N'\!,N}$ are given 
by Eqs.~(\ref{omega-cyc}) and (\ref{f-cyc}).

\subsection{Perturbation corrections}

With account for the coupling $H_3$, the c.m.\ and electronic states and 
radiative transitions are no longer independent. As a result, in contrast 
to the quantum number $\N$, the numbers $s$ and $\nu$ are no longer 
exact ones. Still, when treating the coupling as a perturbation, ion 
states can be designated by the numbers $\N,s,\nu$ enumerating the 
zero-order states. 

Details of the second-order perturbation analysis are given in Appendix~C. 
For the ion energies one obtains 
\be
E_{\N{s}\nu} = E^{\rm Lan}_N(\Omega_{s\nu}) + \E_{s\nu} + \Delta_{s\nu}~. 
\label{ens-result}
\ee
In this equation, $E^{\rm Lan}_N(\Omega_{s\nu})$ is the 
{\em ion Landau energy} modified by the coupling to the internal structure, 
$\E_{s\nu}$ is the zero-order internal energy, and $\Delta_{s\nu}$ is 
the energy shift of the internal states resulting from the 
coupling to the c.m.\ The ion Landau energies are determined by the 
{\em effective cyclotron energies} 
\be
\Omega_{s\nu}=|Q|B/M_{s\nu}~,
\label{Omega_eff}
\ee
where $M_{s\nu}$ is the {\em ion effective mass} dependent on the internal 
states. The explicit expressions for $M_{s\nu}$ and $\Delta_{s\nu}$ are 
given by Eqs.~(\ref{mass-ratio}) and (\ref{shift}). 
 
The perturbation-corrected transition energies and oscillator strengths 
for the bound-ion cyclotron transitions differ from those for the reference ion 
according to the relations
\begin{align}
\omega_{\N'\!,\,\N}& = \lambda_{s\nu}\,\omega^{\rm cyc}_{N'\!,N}~,&
f_{\N'\!,\,\N}& = \lambda_{s\nu}^3 f^{\rm cyc}_{N'\!,N}~,
\label{cyc-second}
\end{align}
where $\lambda_{s\nu} = \Omega_{s\nu}/\Omega = M/M_{s\nu}$. This is the main 
analytical result of our studies. The effect of the internal structure on the 
cyclotron transitions of a complex ion is described by the single parameter 
$\lambda_{s\nu}$. 
Note that in the perturbation regime the energies of a complex ion depend 
linearly on the ion Landau level number, similar to the bare ion. 
To extend the analysis of the ion motion and cyclotron 
transitions beyond the perturbation regime, we apply the coupled-channel 
approach described below. 

\section{Coupled-channel approach}

Numerical multi-electron quantum calculations of the states and radiative 
transitions for a many-particle charged system moving as a whole in a magnetic 
field are yet hardly feasible. We develop a numerical approach applicable to ions 
in which an outer electron is orbiting a ``strongly bound core'', such as the 
nucleus in a one-electron atomic ion or the neutral atom (or a cluster) in a 
multi-electron negative ion. In such ions the interaction between the outer 
electron and the core emerges as a dominant local long-range interaction. 
For positive ions, it is a Coulomb attraction $V(r) = -Z_{\rm eff}/r$, where 
$Z_{\rm eff}$ is an effective charge number for the core ($Z_{\rm eff}$ coincides 
with the nucleus charge number $Z$ for the hydrogen-like ions), and $r$ is the 
separation between the electron and the core. For negative singly-charged ions, 
the dominant long-range interaction is $V(r) = -\kappa/(2r^4)$, where $\kappa$ 
is the polarizability of the neutral core. 

The quantum states of a moving ion are calculated as {\em two-particle states} for 
the motion of the outer electron and the core in a magnetic field. The location of 
the electron is specified by the coordinate {\bf r} with respect to the location 
{\bf R} of the center-of-mass of the core, and the canonical pairs 
\{{\bf R},{\bf P}\} and \{{\bf r},{\bf p}\} of the coordinates and momenta describe 
the motion of the system. 

The c.m.\ motion of the ion along the magnetic field can be separated out. 
The ion states are then described by the wave functions 
$\psi({\bf R}_\perp,{\bf r}_\perp,z)$ in five coupled degrees of freedom. 
A form of the corresponding Hamiltonian, convenient for the numerical treatments, 
is obtained making use of a unitary gauge transformation determined by the operator 
$\exp \left [ ({\rm i}/2)\left( {\bf B} \times {\bf r}_\perp \right) {\bf R}_\perp \right]$, 
see, e.g., Ref.~\cite{BSC2007}. This Hamiltonian is given by the sum of kinetic 
energies of the ion core and the outer electron, and the interaction potential $V$,
\be
H = \frac{K_{0\perp}^2}{2M_0} + \frac{\pi_\perp^2}{2} + V, 
\label{H_cc}
\ee
where $M_0$ is the mass of the core. 

The operators $K_{0\perp}^2 = K_{0x}^2 + K_{0y}^2$ and 
${\mathit \pi}_\perp^2 = {\mathit \pi}_x^2 + {\mathit \pi}_y^2$ are the squares of 
transverse kinetic momenta of the ion core and the outer electron, respectively. 
The electronic kinetic momentum is 
\be
({\mathit \pi}_x,{\mathit \pi}_y,0) = \vv{p}_\perp + (1/2)\vv{B}\times\vv{r}_\perp~, 
\label{pi}
\ee
and the core kinetic momentum has the components 
\be
K_{0x} = {\mathit \Pi}_x - k_x, \;\;\; K_{0y} = {\mathit \Pi}_y - k_y~, 
\label{K0}
\ee
which couple the kinetic momentum~(\ref{Pi}) of the reference bare ion and the 
electronic pseudo-momentum 
\be
(k_x,k_y,0) = \vv{p}_\perp - (1/2)\vv{B}\times\vv{r}_\perp~.
\label{k}
\ee
Notice that $K_{0x}$ and $K_{0y}$ do not commute for positive ions 
which have a positively-charged core, whereas they commute for negative singly-charged 
anions where a core is neutral. 

The integral of motion $K_\perp^2$ for the Hamiltonian~(\ref{H_cc}) is determined by 
the one-particle operator~(\ref{K}) for the motion of the core. The integral of motion 
$\Lt_z$ is the sum~(\ref{Lz-ion}) of the longitudinal angular momenta of the core $L_z$ 
and the outer electron $l_z$. 
 
To compute the wave function $\psi({\bf R}_\perp,{\bf r}_\perp,z)$, we employ a 
coupled-channel approach~\cite{Friedrich_book}. We introduce {\em two-particle} basis 
functions, dependent on ${\bf R}_\perp$ and ${\bf r}_\perp$, as the 
{\em common eigenfunctions} of $K_\perp^2$ and ${\cal L}_z$. The basis states and the 
computed quantum states of ion are thereby attributed to the quantum numbers 
$N_0$ and $\Lt$. The wave function of the ion is expanded in the basis set (channels), 
with the expansion coefficients being the functions of the remaining degree of 
freedom, $z$. These functions, along with the quantum energies, are found by solving 
a set of coupled second-order differential equations. 

Mathematical details of the coupled-channel calculations are given in Appendix~D. 
Similar to the perturbation treatment, the calculated ion energies depend on 
the sum ${\cal N} = N_0 + {\cal L}$, which enumerates the states of the 
quantized collective motion degenerate with respect to $N_0$. 
For each $\N$, the calculations yield a series of bound states 
with different properties of the internal motions. It is convenient to enumerate 
these states by the integer $s=0,1,2,\ldots$ according to ascending 
order of the energies $E_{\N{s}}$: $E_{\N{0}}<E_{\N{1}}<E_{\N{2}}<\ldots$. 
For each $s$, the energies increase with increasing $\N$ 
forming ``$s$-branches'' of the levels. 
Computing the ion states as the eigenstates of $K_\perp^2$ and ${\cal L}_z$ 
results in the following possibilities to vary the number $\N$ for 
a given $s$-branch: $\N=s,s+1,s+2,\ldots$ for positive ions, and 
$\N=-s,-s+1,-s+2,\ldots$ for negative ions. For an infinitely heavy ion, 
the energies $E_{\N{s}}$ coincide with the electronic energies and become 
degenerate with respect to $\N$, 
\be
E_{\N{s}} \to \es \;\;\; \mbox{at} \;\;\; M \to \infty~.
\label{correspondence}
\ee
Once the quantum states of the ion are computed, the multi-channel wave functions 
are used to compute the bound-ion cyclotron transitions with the selection 
rules~(\ref{bi-cyc}). 

\section{Numerical studies}

For numerical calculations of the bound-ion cyclotron transitions we have 
selected two particular examples of ions. 

One example is He$^+$, the lightest in the group of hydrogen-like (one-electron) 
ions. We consider the lowest (tightly-bound) internal states~\cite{BezPavVen98} 
which appear in magnetic fields $B \gg Z^2$, where $Z$ is the nucleus charge 
number, and the atomic units are used for the magnetic field strength. For these 
states of He$^+$, the coupling to the c.m.\ becomes prominent in extremely strong 
magnetic fields. We compute the bound-ion cyclotron transitions for magnetic fields 
of $10^8-10^9$~T, typical for the neutron stars. The calculations are particularly 
motivated by the expectation that the ions He$^+$ can play an important role in 
the atmospheres of neutron stars~\cite{PB2005}. 

Another example is the atomic and cluster negative ions bound exclusively due to 
the presence of the magnetic field, the so called 
{\em magnetically-induced anions}~\cite{BSC-review,BSC2007}. Our studies apply 
to magnetic fields $B \ll 1$ for which the excess electron is bound in a diffuse 
orbital extending well beyond the orbitals of the core electrons. 
The coupling of this electron to the c.m.\ noticeably affects the ion cyclotron 
transitions for magnetic fields of $10-100$~T achievable in terrestrial 
laboratories. Being of general theoretical interest, these studies 
are relevant to possible experiments with the magnetically-bound anions.

The quantum numerical calculations for the selected examples have a lot in common 
and follow the above described coupled-channel approach. The He$^+$ ion is a 
one-electron system. The magnetically-induced negative ions can also be considered 
as one-electron systems in which the excess electron is attached to the neutral 
counterpart. 

We remark that for positive ions, the excess electron can be bound in two 
different classes of states, tightly-bound and hydrogen-like ones. They are 
different with respect to the quantum motion of the electron along the magnetic 
field: the tightly-bound states are the ground states for this motion, whereas 
the hydrogen-like states are excited ones. In contrast to positive ions, 
magnetically-bound negative ions typically exist only in the ground states for 
the longitudinal motion of the excess electron (see, e.g., Ref.~\cite{BSC2000}). 
The difference stems from the different character of the long-range interaction 
between the outer electron and the ion core (a Coulomb attraction $\propto r^{-1}$ 
for positive ions, and a much weaker polarization attraction $\propto r^{-4}$ for 
negative ions). Our studies do not address the hydrogen-like states for the 
positive ion He$^+$, i.e., only the ground states for the longitudinal motion of the 
bound electron, both for He$^+$ and magnetically-induced anions, are considered. 
The quantum states are therefore all related to the value $\nu=0$ corresponding to 
a nodeless structure of the wave function in the longitudinal coordinate, and we omit 
the unnecessary label $\nu$ in what follows. 

\subsection{Positive ions: He$^{+}$ in strong magnetic fields} 

\begin{figure*}[t]
\centering
\begin{tabular}{cc}
\includegraphics[width=0.435\textwidth]{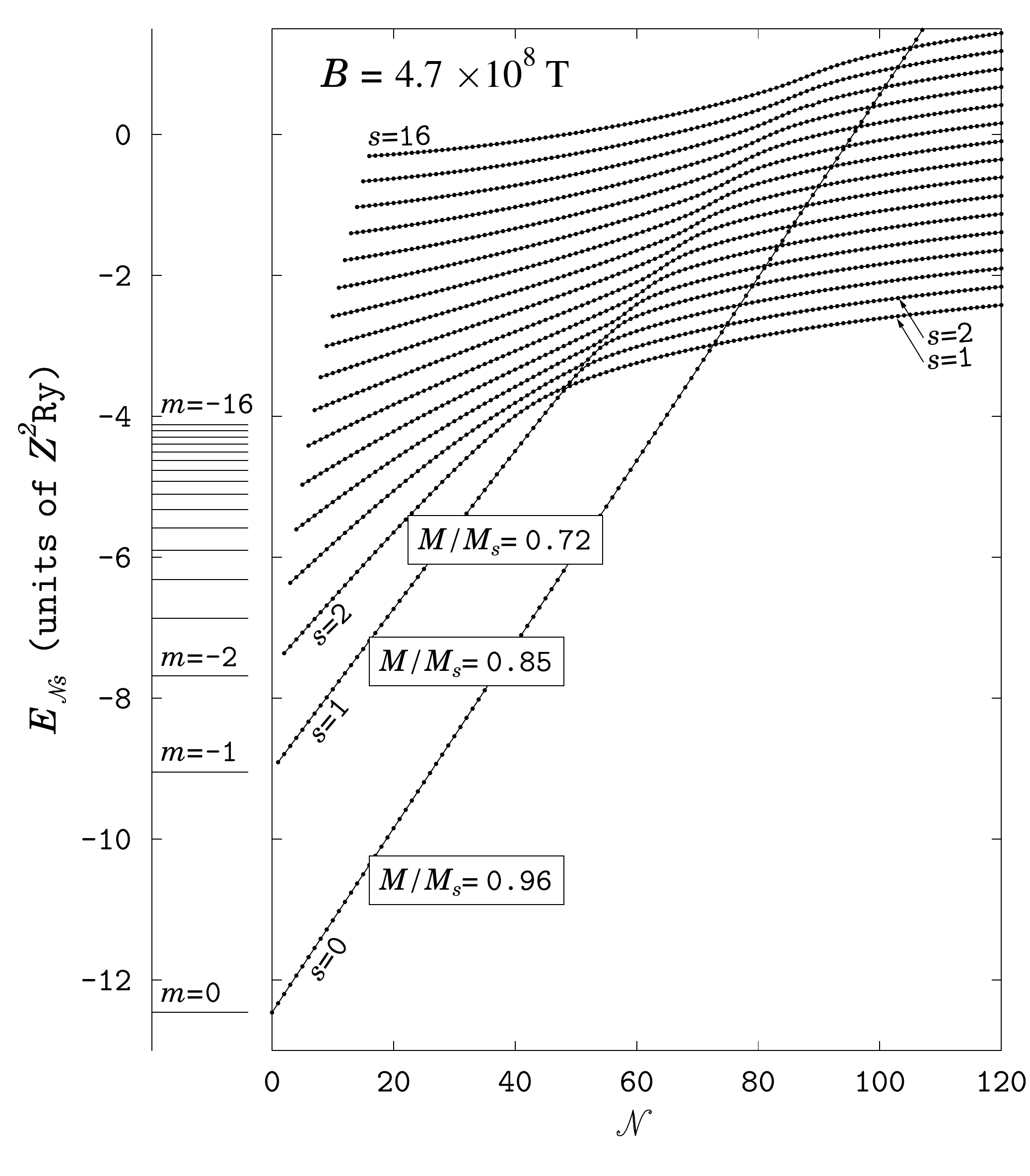} &
\includegraphics[width=0.435\textwidth]{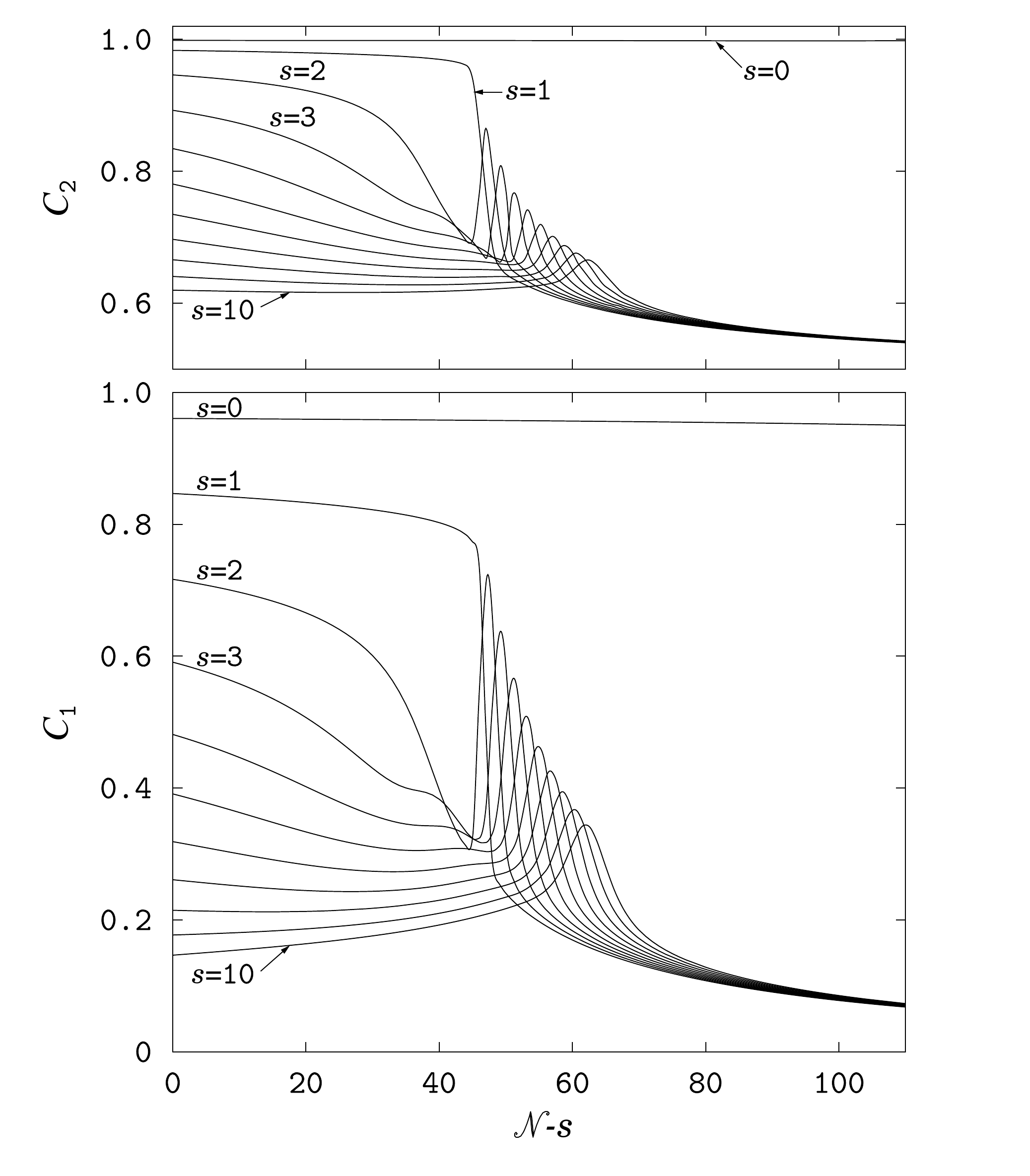} \\
\noalign{\vspace*{-1ex}}
\includegraphics[width=0.435\textwidth]{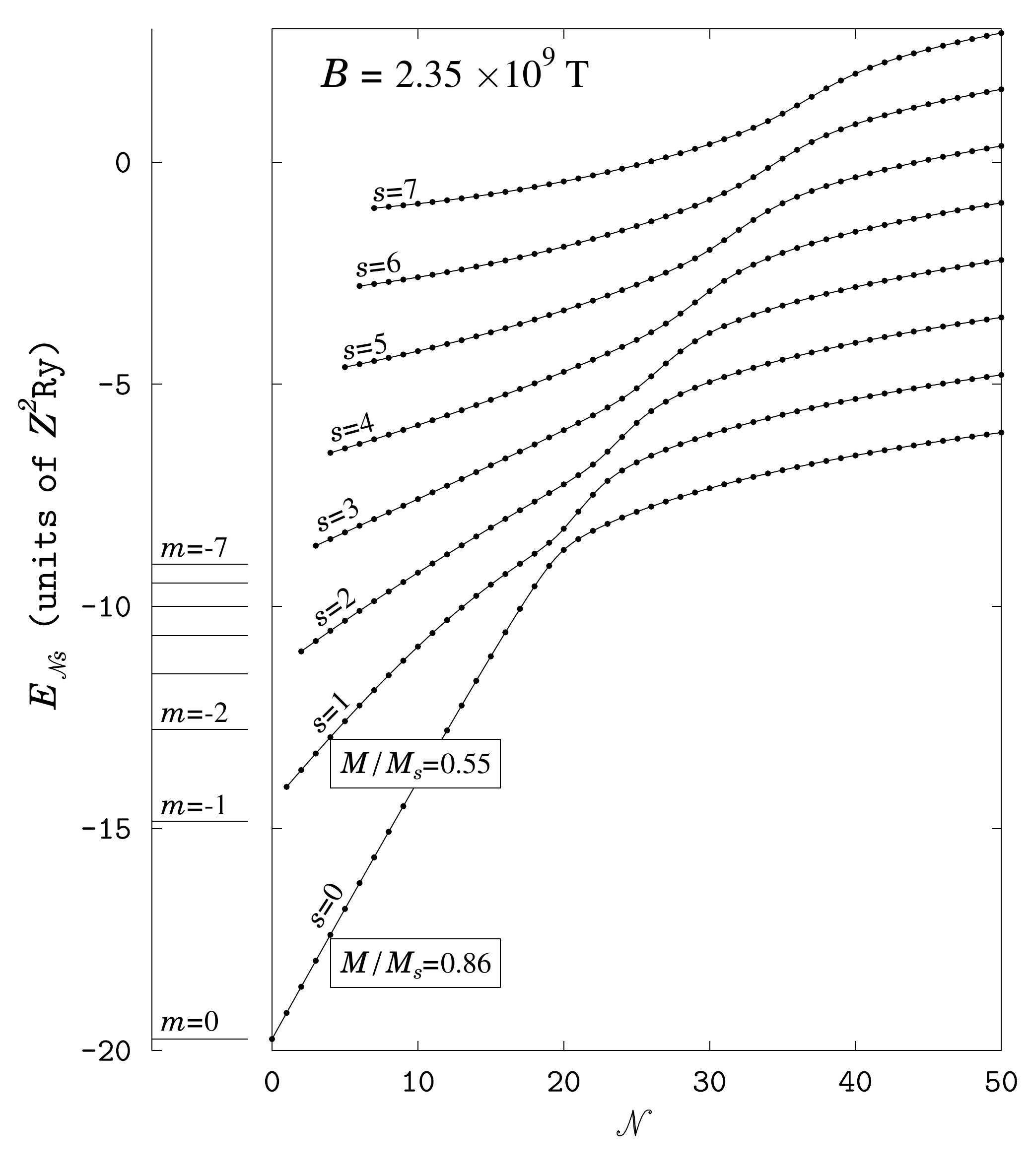} &
\includegraphics[width=0.435\textwidth]{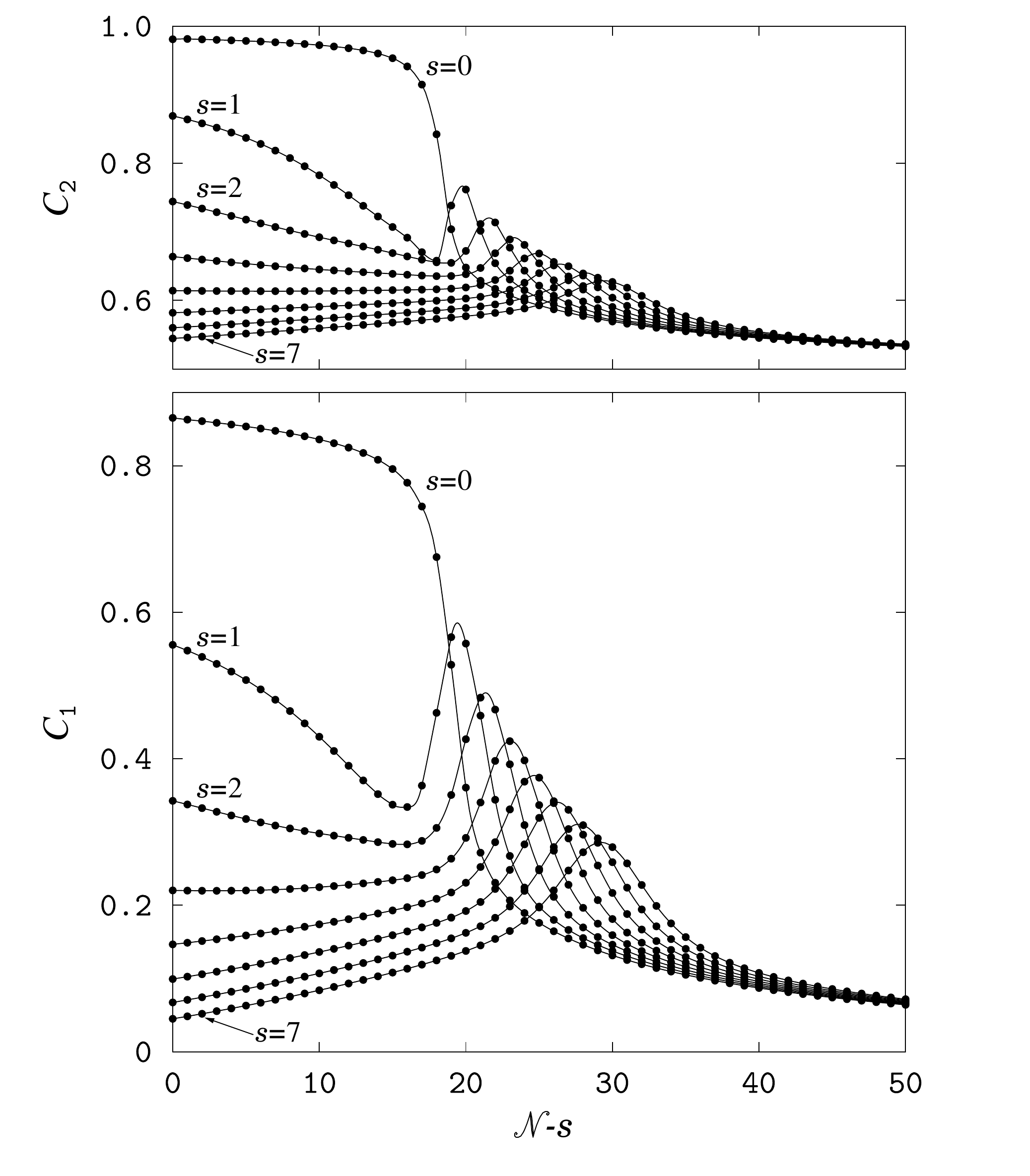}
\end{tabular}
\vspace*{-1ex}
\caption{Energy levels and parameters of the bound-ion cyclotron transitions 
for He$^+$ in the magnetic field of $4.7 \times 10^8$~T (top) and 
$2.35 \times 10^9$~T (bottom). In the left plots dots show the energies of the 
tightly-bound states for different values of the quantum number $\N$. Smooth 
solid lines connect the energies for visualization of the $s$-branches, and the 
ratios $M/M_s$ are deduced from the linear parts of the branches. 
The energies for an infinitely heavy ion are shown by the horizontal bars 
labeled by the magnetic quantum number $m$. 
Right plots show the parameters $C_1$ and $C_2$ (see Eq.~(\ref{C1C2})) 
for the bound-ion cyclotron absorption transitions along the computed 
$s$-branches of states.}
\label{He+}
\end{figure*}

The results of coupled-channel calculations of the quantum states and bound-ion 
cyclotron transitions for the He$^+$ ion moving in a strong magnetic field are 
presented in Fig.~\ref{He+}. 
Since the c.m.\ motion along the magnetic field can be separated out, we exclude 
the corresponding (additive) kinetic energy from the computed ion energies. 
We also exclude the values of the electron and nucleus 
zero-point Landau energies (for the infinitely heavy ion only the electron 
zero-point energy matters). Therefore, the zero energy value is the continuum 
edge for the quantum levels $E_{\N{s}}$ 
of the moving ion (as well as for the levels $\es$ of the 
infinitely heavy ion shown in the figures for the reference). 
Notice that our coupled-channel approach applies to an arbitrary 
hydrogen-like ions. The energies $\es$ 
for infinitely heavy ions scale with the nucleus charge number as 
$\es \propto Z^2$. 
That is why we use the units $Z^2$Ry ($1~\mbox{Ry} = 13.606~\mbox{eV}$) 
for the energies shown in the figures ($Z=2$ for the ion He$^+$). 

The properties of He$^+$ for the magnetic field strength of $4.7 \times 10^8$~T 
(the corresponding value in atomic units is $B = 2000$) are demonstrated in the 
top plots of Fig.~\ref{He+}. The discrete energies of the ion shown by dots 
group into the $s$-branches shown by smooth solid lines. For an infinitely 
heavy ion, the branches merge to the internal electron energies $\es$ 
shown by horizontal bars and labeled by the values of the conventional 
magnetic quantum number $m=-s$, cf. the relation~(\ref{correspondence}). 
These energies have been addressed both analytically and numerically 
in many previous studies (see, e.g., Ref.~\cite{Ruder_1994} and the references 
therein) and are reproduced in our calculations by setting $1/M=0$. 

The property important for identifying the bound-ion cyclotron 
transitions are linear-like dependencies of the energies $E_{\N{s}}$ on $\N$. 
The numerically calculated energies depend on $\N$ almost linearly for the 
entire $s=0$ branch. At higher $s$, the growth of $E_{\N{s}}$ with $\N$ 
is close to linear only for low-lying energies, $\N=s,s+1,s+2,\ldots$. 
The linear parts of the energy branches correspond to the perturbation regime, 
considered in Sec.~III.B. By fitting these parts by the 
formula~(\ref{ens-result}), the ratios $M/M_s$ are estimated and indicated 
in the plot. The values for the ratios are lower for higher $s$-branches: 
$0.96$, $0.85$ and $0.72$ for $s=0$, $1$ and $2$, respectively. Thus, the 
ion's effective masses increase with increasing the internal excitations. 

At $\N\to\infty$ the couplings between the channels vanish, and the quantum 
states converge to the states computed within a single-channel (adiabatic) 
approximation. The ion energies approach the values that are multiples of 
the nucleus cyclotron energy $\Omega_0 = ZB/M_0$, where $M_0$ is the 
nucleus mass. These thresholds correspond to the detached electron and 
nucleus, with the electron occupying the ground Landau level and the 
nucleus occupying the ground as well as the excited levels. 

At intermediate values of $\N$, the neighboring branches approach each other. 
To visualize the branches of the discrete energy levels in the domains of 
nearest approach, the levels can be connected by lines which either cross 
or avoid the crossings. We opted to show the 
line $s=0$ as crossing the other lines, and the lines $s=1,2,\ldots$ as 
exhibiting avoided crossings. Thus, with increasing $\N$, the energies 
$E_{\N{s}}$ increase infinitely for $s=0$ and approach the thresholds 
$(s-1)\Omega_0$ for $s=1,2,\ldots$. 

Notice that the states with $E_{\N{s}}>0$ belong to the 
continuum. Neglecting the coupling of the c.m.\ to the internal motions, 
the positive-energy states are computed as bound, whereas in reality they 
are quasi-bound (auto-ionizing) states because of the coupling. 
All the branches shown in Fig.~\ref{He+} have the minimum-energy $\N=s$ states 
below the continuum $E=0$. The set of such branches for $B = 4.7 \times 10^8$~T 
is restricted to $s=0,1,\ldots,16$. Higher branches comprise the quasi-bound 
states only, and we do not include them in the figure. 

As Fig.~\ref{He+} shows, the levels with $\N = s$ for the bound moving 
He$^+$ are shifted towards the continuum edge with respect to the levels 
with $m=-s$ for the infinitely heavy ion. This implies that already 
the lowest states in the moving ion are bound weaker than the states with the 
similar electronic excitations in the infinitely heavy ion. The only exception 
is the lowest state $\N=s=0$, whose binding energy is not affected by the 
c.m.\ motion. The energy offsets increase with increasing $s$ resulting in the 
finite number of branches below the continuum edge. 

According to the general considerations presented above, the bound-ion 
cyclotron transitions are the transitions between the neighboring states within 
the computed branches. These transitions satisfy the selection 
rules~(\ref{bi-cyc}). The branch numbers $s$ can be regarded as enumerating 
the internal excitations. For the nearly linear parts of the branches, the 
internal excitations correspond to the same effective masses $M_s$ deduced by 
fitting the energies by the perturbation dependence~(\ref{ens-result}). For the 
non-linear parts around and beyond the avoided crossings the ion states still 
can be considered as states of similar ``internal nature'', based on the 
symmetry and ``smoothness'' of the lines connecting the discrete energy values. 

The transition energies and oscillator strengths for the bound-ion cyclotron 
transitions are computed according to Eqs.~(\ref{D-abs-result}). To quantify the 
deviations of the numerical results both from the results for the reference 
bare ion and from the perturbation results~(\ref{cyc-second}), we have 
calculated and plotted in Fig.~\ref{He+} the parameters 
\bea
C_1 &=& \omega_{\N'\!,\,\N}/\omega^{\rm cyc}_{N'\!,N}~,
\nonumber \\
C_2 &=& \left(f_{\N'\!,\,\N}/f^{\rm cyc}_{N'\!,N}\right)
        \left(\omega^{\rm cyc}_{N'\!,N}/\omega_{\N'\!,\,\N}\right)^3~.
\label{C1C2}
\eea
For the bare ion, we have $C_1 = C_2 = 1$. Within the perturbation 
regime for the coupling between c.m.\ and internal motions, we expect 
$C_1 = M/M_s < 1$ and $C_2 = 1$. 

For $B=4.7 \times 10^8$~T, in close relation to the properties of the ion 
energies, the parameters $C_1$ and $C_2$ show the perturbation regime to hold 
for the bound-ion cyclotron transitions along the entire $s=0$ branch as well 
as for the transitions involving the states up to $\N \approx 45$ along the 
$s=1$ branch. For higher internal excitations, $C_1$ and $C_2$ deviate from 
constant values and depend on the quantum number $\N$, which reflects the 
effects of the internal structure on the cyclotron transitions of the entire 
ion. For $s>1$, these effects are already prominent starting from the lowest 
transition states $\N=s$. 

Similar properties of the levels and bound-ion cyclotron transitions are 
displayed by He$^+$ in a higher magnetic field of $2.35 \times 10^9$~T 
($B=10\,000$ in atomic units). The results of numerical calculations are 
presented in the bottom plots of Fig.~\ref{He+}. We have restricted the 
computations to the branches that are below the continuum at $\N =s$. The 
number of these branches, $s=0,1,\ldots,7$, is smaller than the number of 
branches studied for the lower magnetic field strength. The branches are 
well separated from each other, including the domains of avoided crossings. 
Therefore, we have selected the $s=0$ branch as undergoing the avoided 
crossing and approaching the continuum threshold $E=0$ at $\N\to\infty$, 
in contrast to the $s=0$ branch in the top plot of the figure. For higher 
magnetic fields, the perturbation treatment of the coupling of the c.m.\ to 
the electronic motion becomes progressively less accurate. The numerical 
multi-channel calculations show that the ion properties can be described in 
terms of effective masses for only a small portion of the quantum states: 
$\N \lesssim 15$ for the $s=0$ branch and a few lowest states for the $s=1$ 
branch. The corresponding ratios of the mass of the ion to the effective masses 
are $0.86$ and $0.55$, respectively. For higher $s$-branches, the effects of 
the internal structure on the bound-ion cyclotron transitions are essentially 
non-perturbative.

\subsection{Negative ions: atomic and cluster magnetically-induced anions}

As the magnetically-induced anions is a rather unconventional type of ions, 
it is worthwhile to outline their binding mechanism and properties before 
discussing the bound-ion cyclotron transitions. The origin of the binding is 
a combination of the long-range electron correlation in the anion and the 
confining magnetic field. At large distances $r$ from the neutral core to 
the excess electron the correlation results in a local polarization potential 
with a ``strength'' determined by the polarizability $\kappa$ of the 
core\footnote{In general case, a polarizability is anisotropic and given by 
a tensor. We restrict our analysis to the isotropic polarizabilities determined 
by the single values $\kappa$.}. 

In the absence of magnetic field, the correlation and emerging long-range 
potential play a primary role, with respect to the role of non-local 
short-range interaction, in forming the so-called correlation-bound 
anions~\cite{Sommerfeld_2010}. 
An external magnetic field adds a lot to stability of the anionic states and 
makes the long-range binding possible when the polarization potential is too 
weak to support a stable anion without the field. In addition, the presence 
of a magnetic field enables a long-range binding in a sequence of excited 
anionic states. The magnetically-induced anionic states thus emerge as the 
states formed exclusively in the magnetic field.  

For infinitely heavy anions, the sequence of the 
magnetically-induced anionic states was formally predicted to be 
infinite~\cite{AHS_1981}. The states are characterized by the quantized 
values~(\ref{l-vals}) of the longitudinal angular momentum of the attached 
electron. The corresponding electronic energies $\E_s$ have been explicitly 
estimated in Ref.~\cite{BSC2000}:
\begin{align}
\E_0 &= -0.31\kappa^2 B^2~,         & s &= 0~,
\label{binding_0} 
\\
\E_s &= -0.12\kappa^2 B^3 \lambda_s^2~, & s &= 1,2,\ldots~,
\label{binding_s}
\end{align}
where $\lambda_1 = 1$, $\lambda_s = [1-(1.5/s)]\lambda_{s-1}$ for $s \geq 2$, 
and the atomic units are used for the energies, polarizability and the magnetic 
field strength. These estimates apply to small field strengths, $B \ll 1$. 
The energies $\E_s$ do not include the zero-point Landau energy for 
the electron, so that $|\E_s| = -\E_s$ are the binding energies. 
Different scaling of the energies with the 
the field strengths for the ground $s=0$ and excited $s>0$ states results from 
different localization of the excess electron in the plane transverse to the 
field. For $s=0$ the electron density is maximal at the location of neutral 
system, whereas for $s>0$ the maximum of the density is shifted away from the 
system. As a result, $\E_0$ is sensitive to a divergent behavior 
of the polarization potential at the origin, and the numerical prefactor in 
Eq.~(\ref{binding_0}) is model-dependent. For details, we refer to the 
studies~\cite{BSC-review,BSC2007}. In contrast, the energies of excited states 
are not sensitive to the core of polarization potential in the presence of the 
field. It has also been recognized~\cite{BSC-review,Dubin} that the $s=0$ state 
manifests itself only for systems which do not form stable anions without 
magnetic field. The sequence $s=1,2,\ldots$ of excited states emerges therefore 
on top of the ground magnetically-induced $s=0$ state or of the conventional 
anionic state perturbed by the magnetic field. 

With account for the c.m.\ degrees of freedom, the energies $E_{\N{s}}$ of the 
magnetically-induced states are specified, in addition to $s$, by the quantum 
number $\N$. The coupling between the electronic and c.m.\ motions destroys the 
states with high $s$, and the sequence of magnetically-induced states turns out 
to be finite~\cite{BSC-review}. Depending on systems and on magnetic field 
strength, there can be a few $s$-branches of the bound states, a single 
branch or even a single state or no bound states at all.  

\begin{table*}[ht]
\caption{
The cyclotron energies $\Omega$ of the bare ions, the energies $\E_s$ of the 
infinitely heavy anions, and the parameters of magnetically-induced energy branches 
$E_{\N{s}}$ of moving anions (see Eq.~(\ref{cc-energies})) at the magnetic field 
strengths of $50$~T. The number $s$ numerates the branches of bound states, 
$E_{\N{s}} < 0$, for the moving anions. The numbers of branches are finite and 
restricted to those studied in the table. 
}
\label{Anions}
\setlength{\tabcolsep}{5pt}
\begin{tabular}{c r@{.}l c r@{.}l r@{.}l r@{.}l r@{.}l r@{.}l}
\noalign{\medskip}
ion & 
\multicolumn{2}{c}{$\Omega$ (MHz)} &
$\;s\;$ & 
\multicolumn{2}{c}{$-\E_s$ (MHz)} & 
\multicolumn{2}{c}{$-E_{-s,s}$ (MHz) $^{\rm a)}$} & 
\multicolumn{2}{c}{$\Omega_s$ (MHz) $^{\rm a)}$} & 
\multicolumn{2}{c}{$M/M_s$ $^{\rm a)}$} &
\multicolumn{2}{c}{$M/M_s$ $^{\rm b)}$} \\ 
\noalign{\smallskip}\hline\noalign{\smallskip}
Xe$^-$      &  5&800               & 0 &   7&135$\times 10^4$ &  7&134$\times 10^4$  &  5&800   & 1&     & 1&     \\
\noalign{\smallskip}
Xe$_4^-$    &  1&447               & 0 &   1&142$\times 10^6$ &  1&142$\times 10^6$  &  1&447   & 1&     & 1&     \\
            & \multicolumn{2}{c}{} & 1 &  94&02               & 91&17                &  1&402   & 0&9685 & 0&9796 \\
            & \multicolumn{2}{c}{} & 2 &   5&876              &  2&452               &  0&5046  & 0&3486 & 0&4918 \\
\noalign{\smallskip}
Xe$_{13}^-$ & 0&4462               & 1 & 993&1                & 992&2                &  0&4458  & 0&9991 & 0&9994 \\
            & \multicolumn{2}{c}{} & 2 &  62&07               &  60&74               &  0&4338  & 0&9722 & 0&9755 \\
            & \multicolumn{2}{c}{} & 3 &  15&52               &  13&81               &  0&3800  & 0&8516 & 0&8618 \\
            & \multicolumn{2}{c}{} & 4 &   6&062              &   4&084              &  0&2597  & 0&5820 & 0&6207 \\
            & \multicolumn{2}{c}{} & 5 &   2&970              &   0&9017             &  0&09799 & 0&2196 & 0&3461 \\
\noalign{\smallskip}\hline\noalign{\smallskip}
Ar$^-$      & 19&06                & 0 &   1&130$\times 10^4$ &   1&128$\times 10^4$ & 19&04    & 0&9987 & 1&     \\
\noalign{\smallskip}
Ar$_4^-$    &  4&765               & 0 &   1&808$\times 10^5$ &   1&808$\times 10^5$ &  4&765   & 1&     & 1&     \\
            & \multicolumn{2}{c}{} & 1 &  14&893              &   7&293              &  2&272   & 0&4767 & 0&7401 \\
\noalign{\smallskip}
Ar$_{13}^-$ &  1&466               & 1 & 157&3                & 154&4                &  1&438   & 0&9808 & 0&9875 \\
            & \multicolumn{2}{c}{} & 2 &   9&831              &   6&060              &  0&8001  & 0&5456 & 0&6331 \\
\noalign{\smallskip}\hline\noalign{\smallskip}
\multicolumn{12}{l}{$^{\rm a)}$ coupled-channel calculations} \\
\multicolumn{12}{l}{$^{\rm b)}$ perturbation estimates} 
\end{tabular}
\end{table*}

For magnetic field strengths achievable in laboratories, the conventional 
anionic states, if exist, are not much influenced by the field. There, the coupling 
between the excess electron and c.m.\ motion has only a minor if not negligible 
impact on the ion motion and radiative transitions. The examples considered below 
display the same properties for the ground magnetically-induced states. 
In contrast, for the exited states bound due to the field, the coupling 
significantly affects the internal and collective motions, and the bound-ion 
cyclotron transitions differ from those of the reference bare ions. 

For the numerical studies, we have selected the noble gas atoms Xe and Ar and their 
small clusters at a large magnetic field strength $B=50$~T which can be approached 
with resistive magnets in advanced experimental setups~\cite{lab1,lab2}. 
The calculations utilize the polarizability values $27.815$~a.u. for Xe and 
$11.07$~a.u. for Ar provided in the dipole polarizability tables~\cite{dp-tables} 
according to the experimental measurements. 

The closed-shell Xe and Ar do not form conventional stable anions, and thus they 
can display the ground magnetically-bound anionic states. The clusters of these 
atoms can form anions in the absence of magnetic field, due to increasing 
polarization attraction of the excess electron with increasing number of the atoms 
in clusters. However, a minimum number of cluster atoms required to attach the 
electron is known reliably only for the Xe clusters studied by an 
{\em ab initio} Green functions method~\cite{Xe_clusters} and found to form 
stable correlation-bound anions starting from a size of five atoms. 
To address the magnetically-induced states with $s=0$, we have performed 
computations for Xe$_4^-$, which is the largest Xe cluster anion possessing 
this branch of states. To study the effect of cluster size on the energies and 
transitions for the $s>0$ states, we have performed computations for 
Xe$_{13}^-$ formed by a smallest magic number of cluster 
atoms\footnote{The magic numbers correspond to the maximal binding energies 
per atom in a sequence of clusters with the growing number of atoms, see, e.g., 
Ref.~\cite{Cluster_growing} on clusters of noble gas atoms.}. For Ar, we have 
considered the magnetically-induced anions with the same numbers of cluster atoms, 
$4$ and $13$, as for Xe. Since the polarizability of Ar is smaller than that of Xe, 
a minimum number of atoms in a stable conventional Ar cluster anion is larger 
than five. Therefore, Ar$_4^-$ is not stable in the absence of magnetic field, 
and it can display the $s=0$ magnetically-induced states. For Ar$_{13}^-$, we did 
not address the $s=0$ states, for an analogy of the consideration with that for 
Xe$_{13}^-$. 

To compute the magnetically-induced cluster anions, we employ the same 
frameworks of perturbation and coupled-channel approaches as for the atomic 
ions. Since the excess electron resides on a diffuse orbital extending far away 
from the cluster cores, to a good approximation the clusters can be considered 
as entities characterized by masses and polarizabilities. The cluster 
polarizabilities are evaluated as the multiples of atomic polarizabilities and 
the numbers of the cluster atoms. To justify the latter approximation, we 
refer to {\em ab initio} studies~\cite{Xe_dimer} of the polarizability of 
Xe dimer. It was found that though the polarizability of the dimer is anisotropic 
and deviates from the sum of atomic polarizabilities, this does not influence 
significantly the binding energies of the anionic states induced by a 
magnetic field aligned along the dimer axis. The structures of clusters formed by 
$4$ and $13$ atoms (a regular tetrahedron and a regular icosahedron with an extra 
atom at the center, respectively) have higher symmetries, and we may expect even 
a less pronounced, than for the dimer, impact of a non-additivity and anisotropy 
of the cluster polarizabilities on the magnetically-bound anionic states.  

\begin{figure*}[ht]
\centering
\begin{tabular}{cc}
\includegraphics[width=0.43\textwidth]{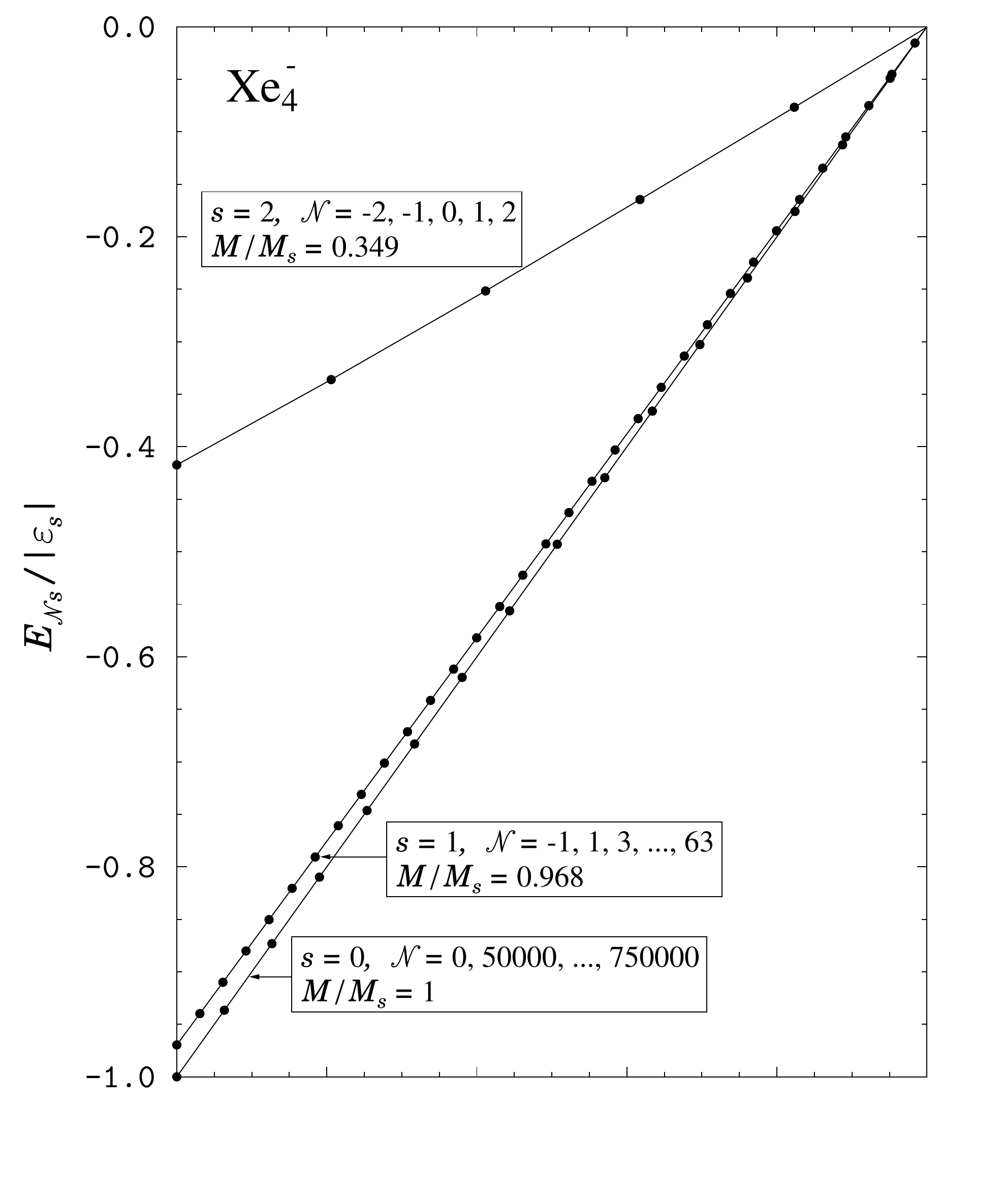} &
\includegraphics[width=0.43\textwidth]{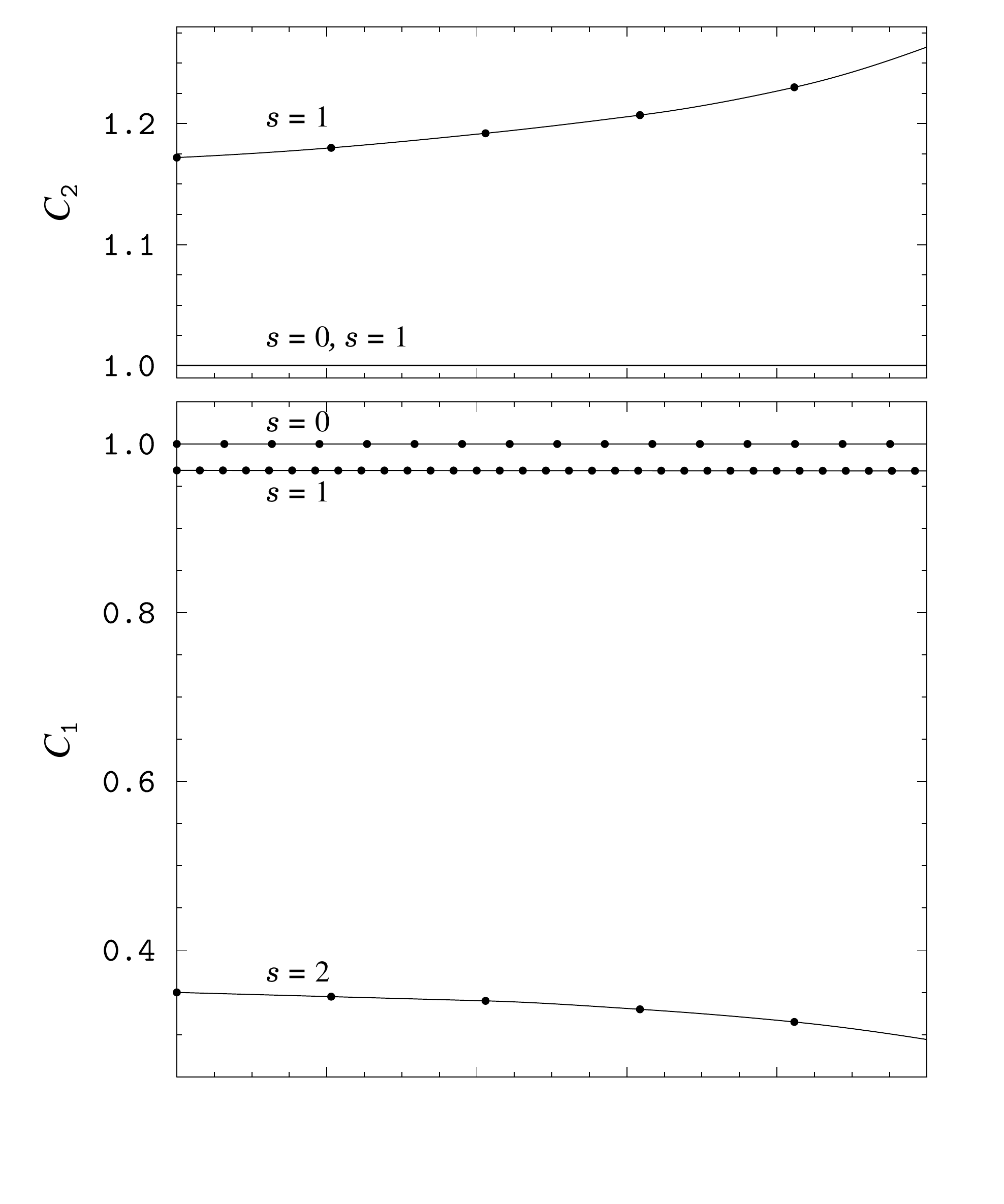} \\
\noalign{\vspace*{-5ex}}
\includegraphics[width=0.43\textwidth]{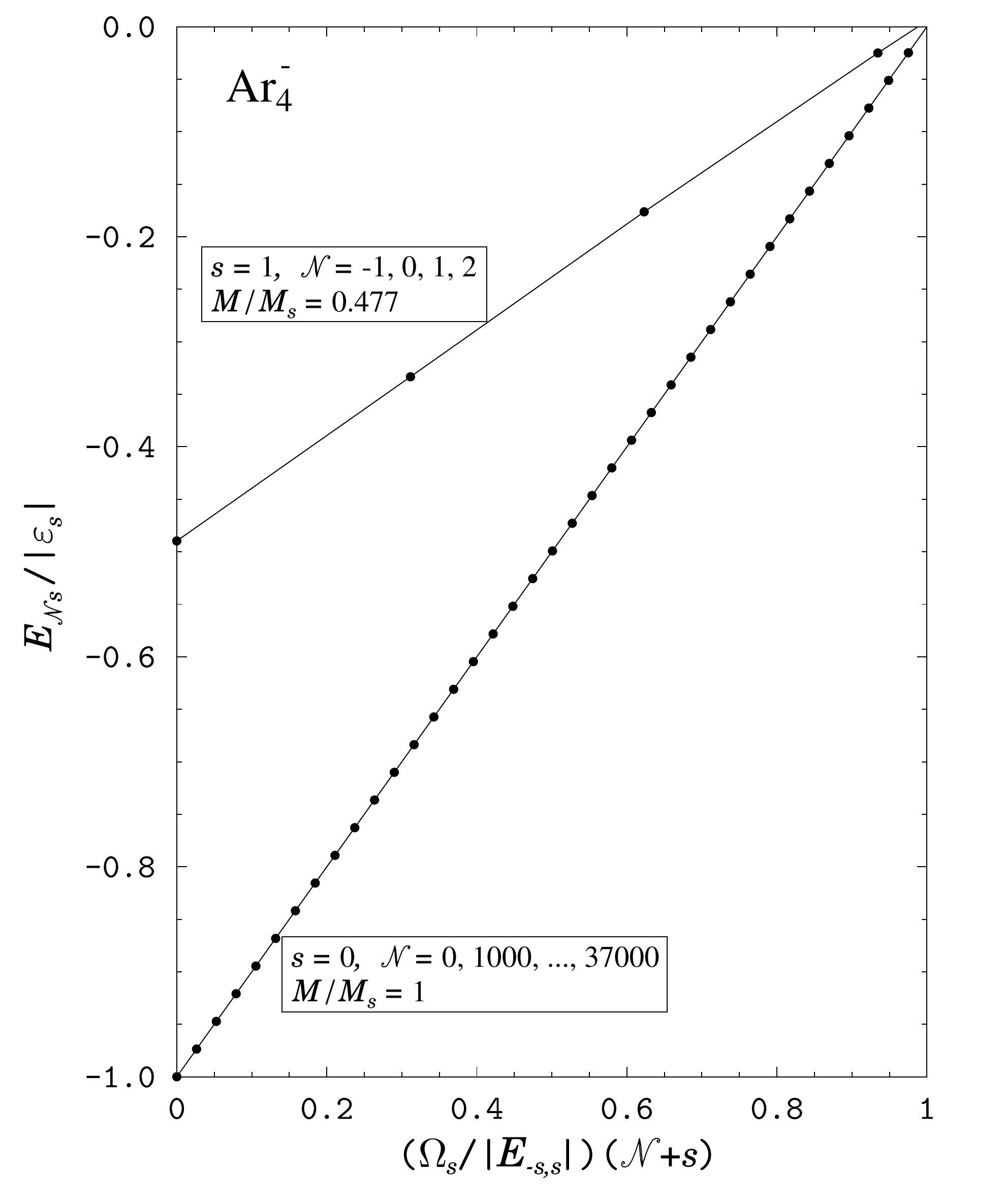} &
\includegraphics[width=0.43\textwidth]{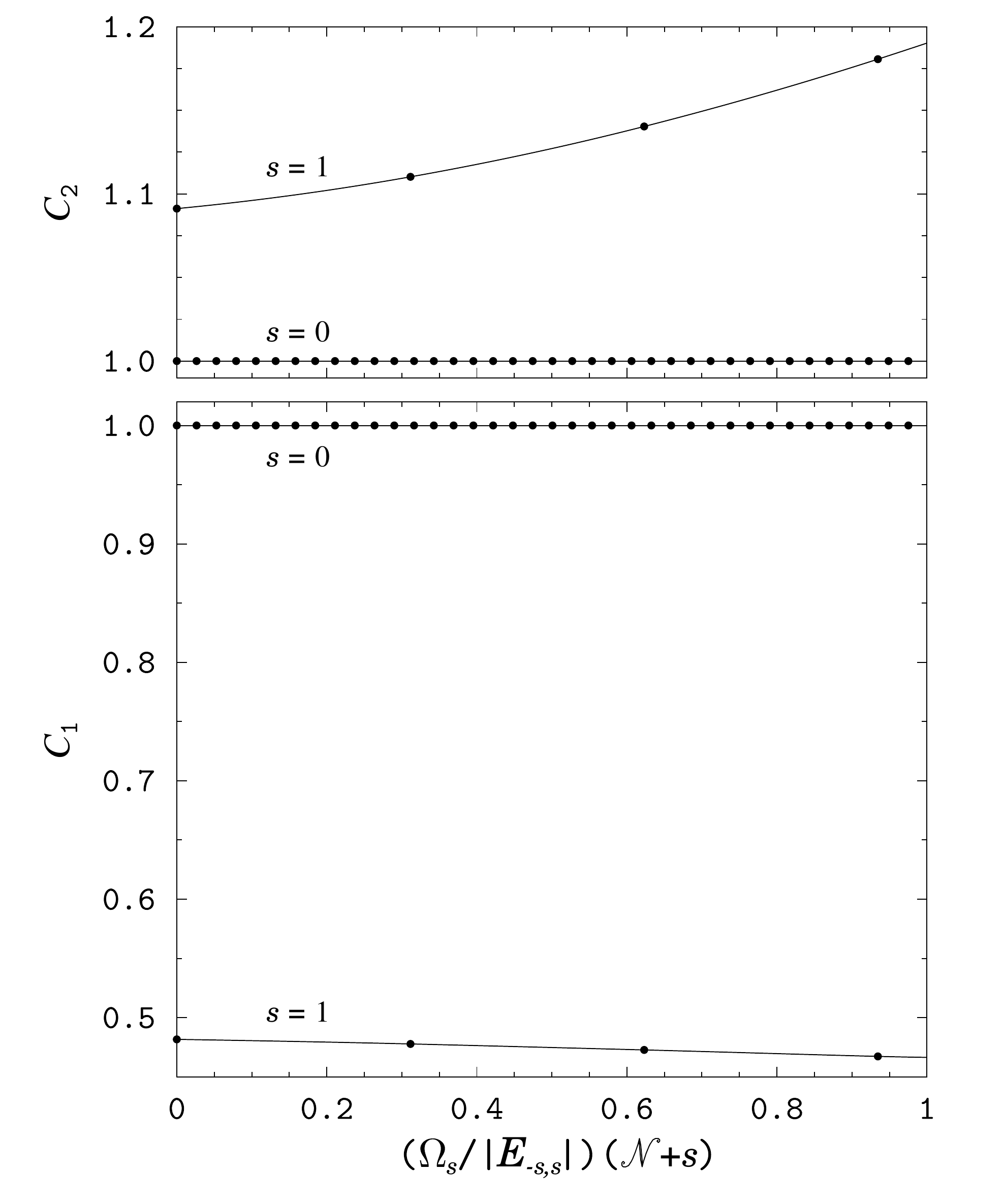}
\end{tabular}
\caption{Energy levels and parameters of the bound-ion cyclotron 
transitions of the magnetically-induced Xe$_4^-$ (top) and 
Ar$_4^-$ (bottom) in a magnetic field of $50$~T. 
The left plots show the $s$-branches of bound levels, with dots displaying 
the discrete levels with the numbers $\N$ as indicated. 
The values $M/M_s$ are determined by approximating the energies by linear 
dependencies on $\N$. The right plots show the 
parameters $C_1$ and $C_2$ of the bound-ion absorption 
cyclotron transitions $\N \to \N+1$ along the branches of the computed levels. 
For Xe$_4^-$, $C_2=1$ for both $s=0$ and $s=1$ as shown by the solid line.}
\label{Xe4Ar4}
\end{figure*}

For the anions considered, the coupled-channel calculations yield the energy levels 
which follow fairly well a linear dependence on $\N$,
\be
E_{\N{s}} = E_{-s,s} + \Omega_s(\N+s)~.
\label{cc-energies}
\ee
This allows one to deduce an effective mass $M_s$ of the ion from the slope $\Omega_s$ 
which has a meaning of the effective cyclotron energy of anion. In Table~\ref{Anions} 
we indicate the values of the minimum energy $E_{-s,s}$, the slope $\Omega_s$ and 
the ratio $M/M_s$ for the $s$-branches computed. We also include the cyclotron energies 
$\Omega$ of the bare ions, the energies $\E_s$ of the infinitely heavy anions, and 
the ratios $M/M_s$ determined according to the perturbation formula~(\ref{mass-ratio}). 

\begin{figure*}[ht]
\centering
\begin{tabular}{cc}
\includegraphics[width=0.43\textwidth]{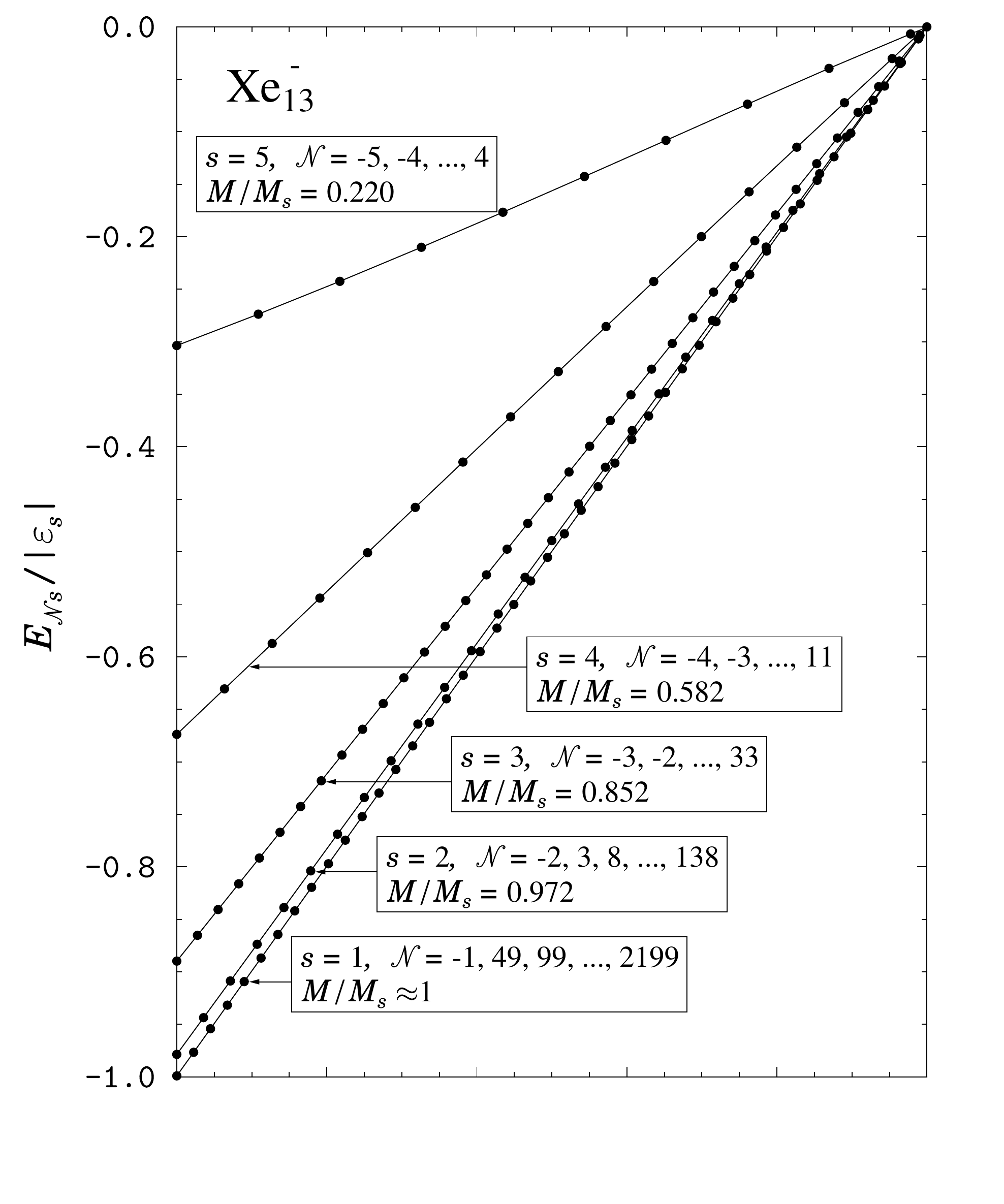} &
\includegraphics[width=0.43\textwidth]{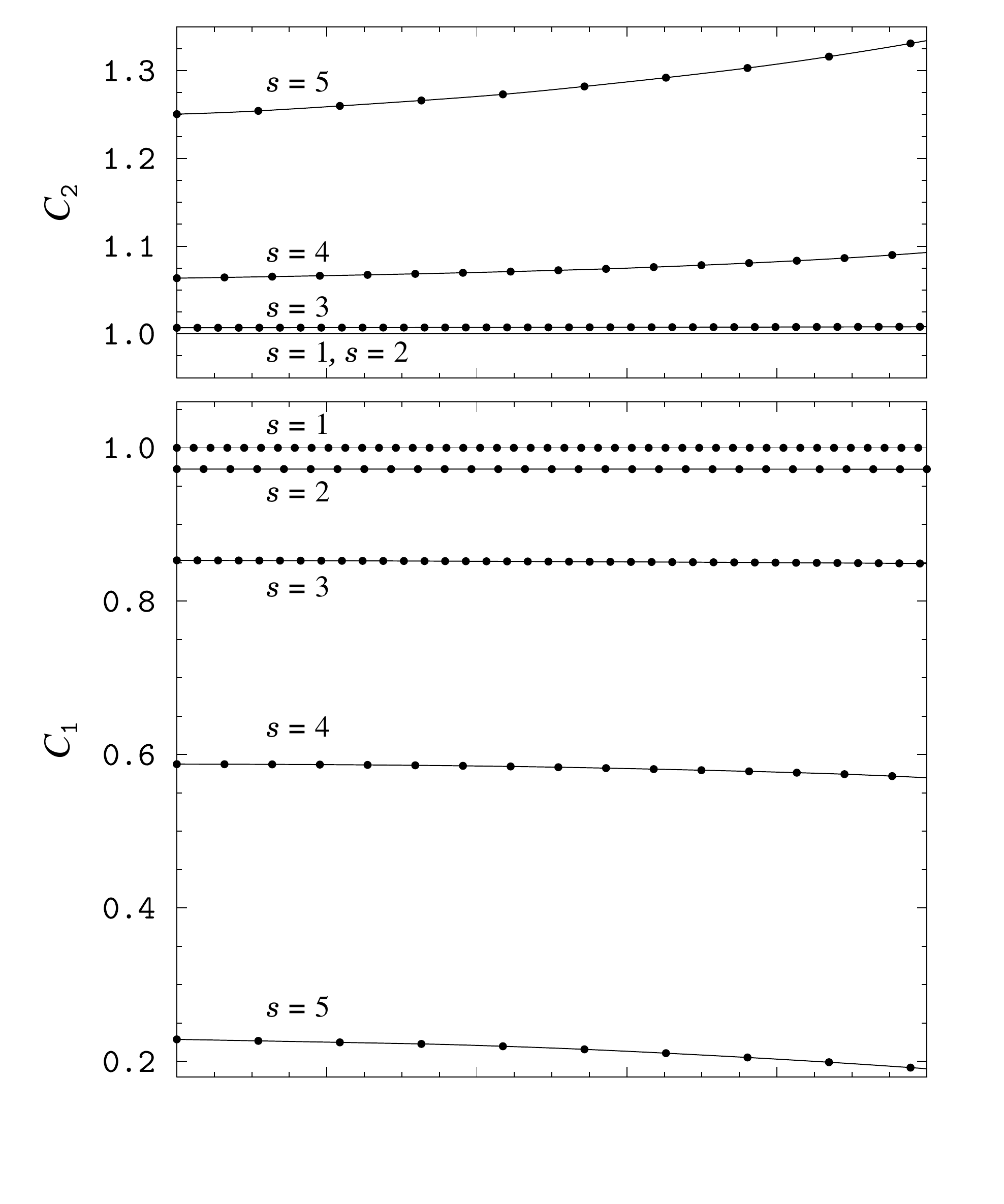} \\
\noalign{\vspace*{-5ex}}
\includegraphics[width=0.43\textwidth]{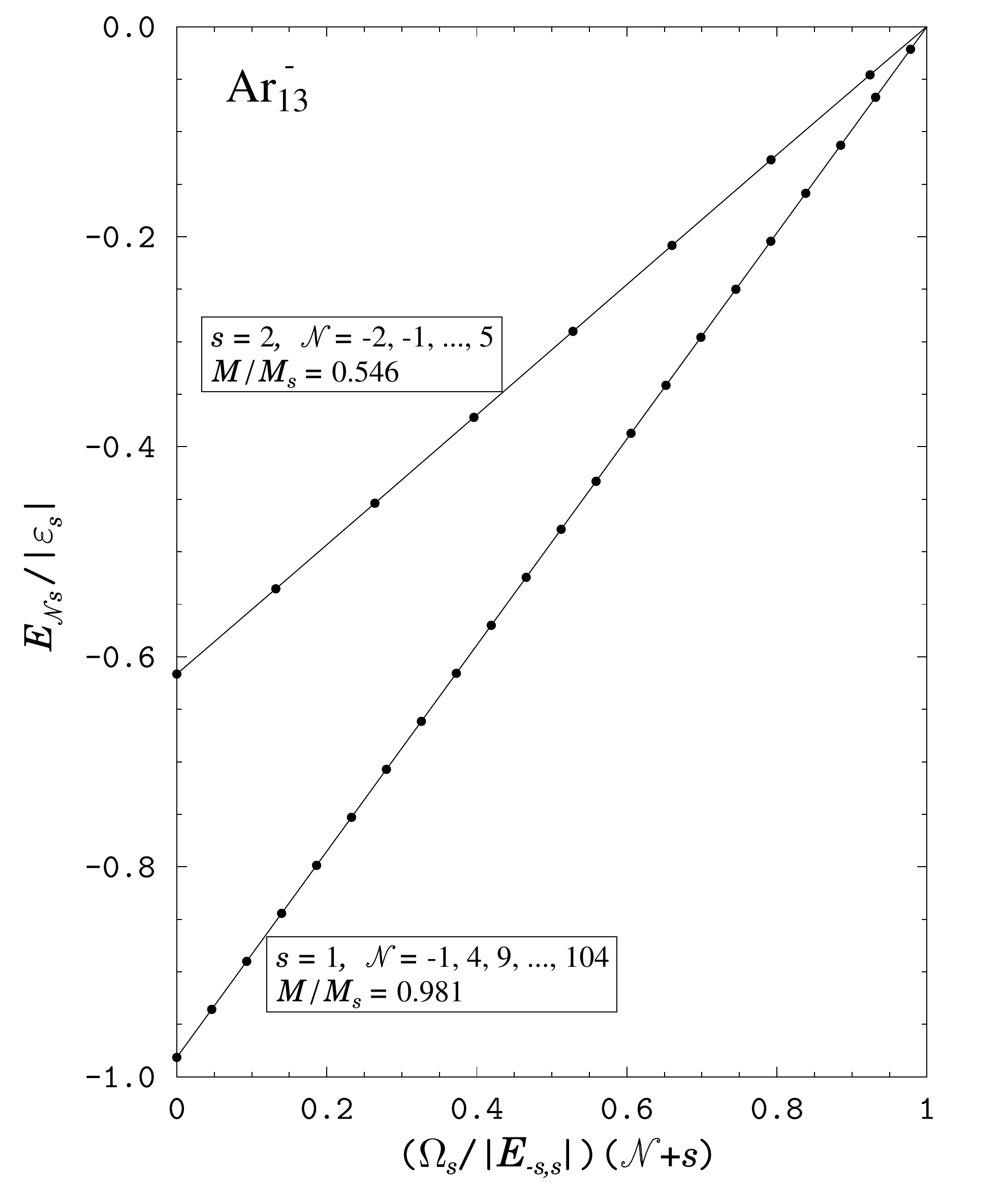} &
\includegraphics[width=0.43\textwidth]{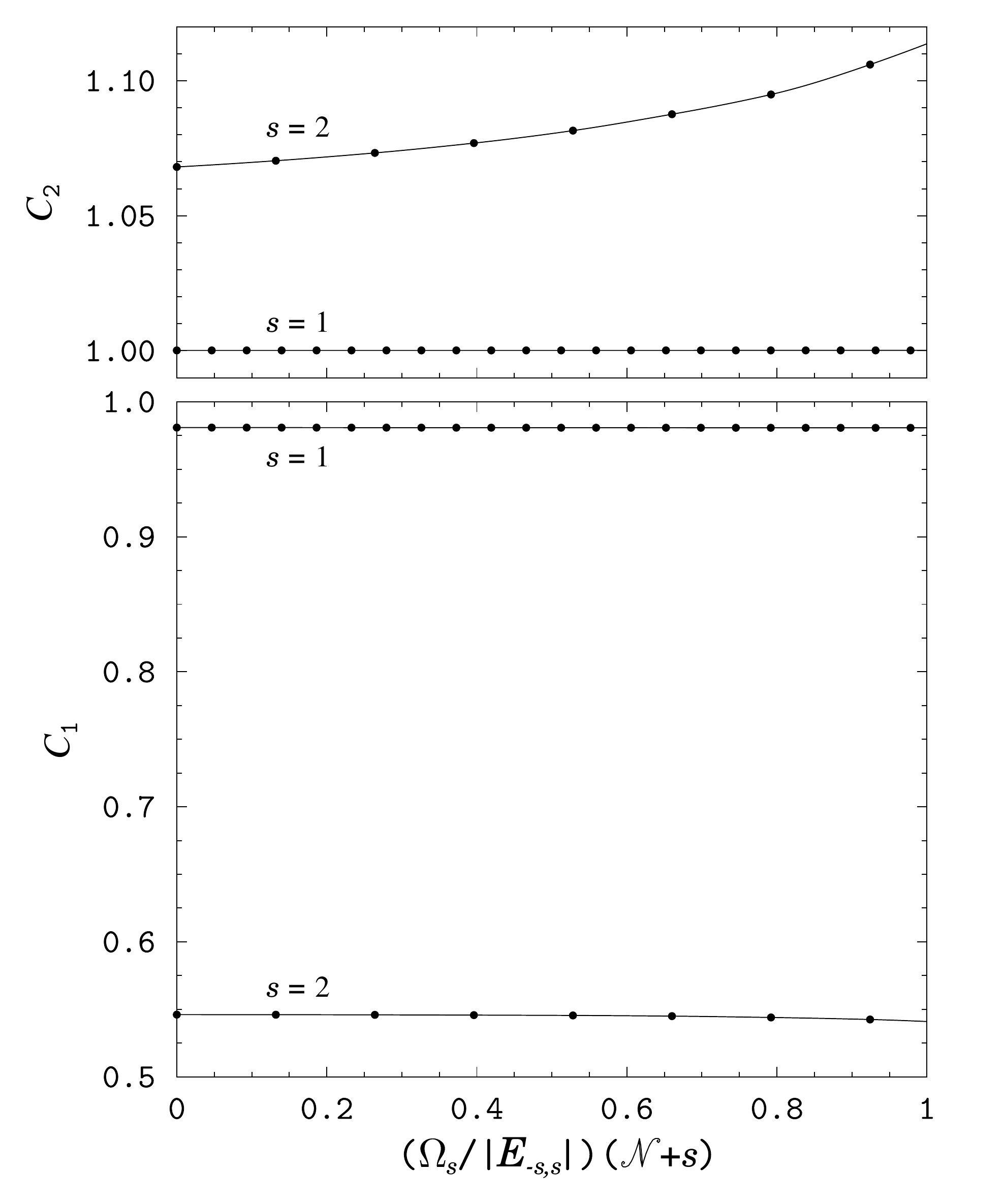}
\end{tabular}
\caption{Same as in Fig.~\ref{Xe4Ar4} but for Xe$_{13}^-$ and Ar$_{13}^-$. 
For Xe$_{13}^-$, $C_2=1$ for both $s=1$ and $s=2$.}
\label{Xe13Ar13}
\end{figure*}

The atomic anions Xe$^-$ and Ar$^-$ display only the $s=0$ branches of the 
magnetically-induced states, whereas the higher $s$-branches are entirely 
destroyed by the coupling between the motions of c.m.\ and excess electron. 
The numbers of the bound states with different $\N$ values are equal to the 
ratios $-E_{-s,s}/\Omega_s$ and are very large in the $s=0$ branches, 
about $12\,300$ for Xe$^-$ and about $600$ for Ar$^-$. The anions move 
essentially as the bare ions ($M/M_s=1$) as found from 
both the coupled-channel and perturbation treatments. Consequently, 
the parameters of the bound-ion cyclotron transitions are $C_1=C_2=1$, 
i.e. the transition energies and the oscillator strengths are the same as for 
the reference bare ions. 

The magnetically-induced anions of Xe and Ar clusters display higher $s$-branches 
of states as well as cyclotron transitions deviating from those of the reference 
bare ions. The results are presented in Figs.~\ref{Xe4Ar4} and \ref{Xe13Ar13}. The 
left plots in the figures show the energies for the anions bound in different states 
of the quantized collective motion. The zero-point Landau energy for the excess 
electron has been subtracted from the computed energies, and the zero energy value 
in the plots corresponds to the detachment threshold. The energies $E_{\N{s}}$ 
for the moving anions are scaled by the binding energies $|\E_s|$ for the 
infinitely heavy anions and shown as functions of $(\Omega_s/|E_{-s,s}|)(\N+s)$, 
where $E_{-s,s}$ is the minimum energy value in the sequence $E_{\N{s}}$ with 
$\N=-s,-s+1,\ldots$, and $\Omega_s$ is the 
effective cyclotron energy determined from the linear growth of the energies with 
$\N$. The scaling is convenient for presenting the results for anionic 
states with significantly different energy values (see Table~\ref{Anions}). 
The right plots in Figs.~\ref{Xe4Ar4} and \ref{Xe13Ar13} show the parameters 
$C_1$ and $C_2$ for the bound-ion cyclotron transitions along the branches of 
levels. 

Figs.~\ref{Xe4Ar4} and \ref{Xe13Ar13} clearly display a linear increase 
of quantum energies with $\N$ increasing along the $s$-branches. The anion's 
motion in the magnetic field is therefore described fairly well in terms of 
the effective masses indicated in Table~\ref{Anions} and in the figures. 
The branches terminate when approaching, with increasing $\N$, the detachment 
threshold $E=0$. Above the threshold, the branches would include the positive 
energies of auto-detaching states calculating which goes beyond the scope 
of this paper. 

With increasing number $N_{\rm at}$ of cluster atoms, the numbers of states 
with different $\N$ in a branch increase as 
$-E_{-s,s}/\Omega_s \sim -\E_s/\Omega \propto N_{\rm at}^3$, since 
$-\E_s \propto \kappa^2 \propto N_{\rm at}^2$, and $\Omega \propto N_{\rm at}^{-1}$ 
due to increasing with $N_{\rm at}$ polarizabilities and masses of the clusters. 
The $s=0$ branches for the cluster anions of four atoms comprise the numbers of 
states exceeding those for the atomic anions by a factor of $64$, 
as reflected in Fig.~\ref{Xe4Ar4} by the numbers of states $\approx 750\,000$ 
for Xe$_4^-$ and $\approx 37\,000$ for Ar$_4^-$. These numbers are very large, 
and the discrete energies in the branches are shown with the increments 
of $50\,000$ and $1\,000$, respectively, in varying $\N$. The numerical 
calculations for the $s=0$ branches yield the effective masses 
coinciding with the masses of anions, and the values $C_1=C_2=1$ for the 
parameters of the cyclotron transitions. As expected from the property 
of the atomic anions, the quantum levels and cyclotron transitions of the 
cluster anions for the $s=0$ branches are not influenced by the coupling between 
the motions of c.m.\ and excess electron. 

In contrast to the atoms, the clusters considered in Fig.~\ref{Xe4Ar4} do 
possess higher $s$-branches of the magnetically-induced anionic states. In 
addition to the $s=0$ branches, the calculations reveal bound states forming 
the $s=1$ and $s=2$ branches for Xe$_4^-$, and the $s=1$ branch for Ar$_4^-$. 
The effective mass exceeds the total mass only slightly for the $s=1$ states 
of Xe$_4^-$ but quite substantially for the $s=2$ states of Xe$_4^-$ and for 
the $s=1$ states of Ar$_4^-$. The values of $C_1$ coincide fairly well with 
the mass ratios $M/M_s$, and the values for $C_2$ do not significantly deviate 
from unity (for the transitions along the $s=1$ branch of levels of 
Xe$_4^-$, the calculations yield $C_2=1$ with a high accuracy). Thus, the effect 
of internal structure on the bound-ion cyclotron transitions follows well the 
perturbation regime.

The larger cluster anions addressed in Fig.~\ref{Xe13Ar13} display more 
magnetically-induced states. For Xe$_{13}^-$, the states group in more branches, 
$s=1,\ldots,5$, than for Xe$_4^-$. For Ar$_{13}^-$, we encounter the same two 
branches, $s=1,2$, as for Ar$_4^-$, which comprise more levels with 
different $\N$. The effective Xe$_{13}^-$ mass for $s=1$ coincides with the total 
mass, and the anion moves in the field essentially as a bare ion. Consequently, 
the bound-ion cyclotron transitions are not affected by the coupling between the 
motions of c.m.\ and excess electron ($C_1 = C_2 = 1$). 
The branches $s=2$ of Xe$_{13}^-$ and $s=1$ of Ar$_{13}^-$ correspond to 
the effective masses only slightly exceeding the total masses, and the cyclotron 
transitions of anions in these internal configurations are only slightly influenced 
by the coupling. For the $s=2$ branch of Xe$_{13}^-$, $C_2=1$ with a high accuracy 
implying an excellent agreement between the perturbation and coupled-channel results. 
For higher $s$ for Xe$_{13}^-$ as well as for $s=2$ for Ar$_{13}^-$, the $C_2$ values 
deviate from unity and slightly increase with increasing $\N$. This implies that the 
perturbation approach becomes progressively less accurate for higher internal 
excitations. 

Overall, the numerical studies demonstrate that the impact of the internal structure 
on the bound-ion cyclotron transitions is guided by the ratio of the binding energy of 
the infinitely heavy ion and the cyclotron energy of the reference bare ion. The smaller 
this ratio is, the more significant are the deviations of the quantized motion of ion
as a whole and the related cyclotron transitions from those of the  bare ion. 
For the magnetically-induced anions considered, these deviations, when emerge, follow 
well a perturbation regime described in terms of the effective masses. 

\section{Conclusions}

We have derived a theoretical framework for quantum description of the motion and 
radiative transitions for ions with internal structure in external magnetic fields. 
We have focused on the collective motion and the related cyclotron-type radiative 
transitions for the entire ions. These {\em bound-ion cyclotron transitions} are 
primarily associated with the properties of the quantum motion of the ions as a whole 
in the field. The coupling of the internal structure of the ions to the collective 
motion in the field makes the transitions to differ from the cyclotron transitions for 
the bare ions with the same masses and charges. The developed theoretical description 
utilizes the integrals of motion for the ions. A general perturbation approach to the 
coupling for the atomic ions is facilitated by the conservation of the longitudinal 
angular momentum for the isolated electronic configuration. The bound-ion cyclotron 
transitions have been identified by specific selection rules. The perturbation approach 
quantifies the transition energies and oscillator strengths for the bound-ion cyclotron 
transitions in terms of the effective masses for the ions. Analytical results have been 
augmented by the results of numerical coupled-channel calculations of the ion states 
and radiative transitions. 

We have studied the bound-ion cyclotron transitions for both positive and negative 
complex ions. Numerical results have been obtained for the He$^+$ ion in strong magnetic 
fields typical for neutron stars, and for the negative magnetically-bound atomic and 
cluster ions for the field strengths typical for laboratory experiments. The calculations 
reveal significant effects of the internal structure on the cyclotron transitions of the 
entire ions. For the He$^+$ ions, the presented theoretical studies could be used for 
interpretations of the spectral observations of neutron stars with strong magnetic fields. 
The negative magnetically-induced ions are yet to be directly probed in an experiment 
on magnetically-guided electron attachment. In particular, experimental detection of the 
bound-ion cyclotron transitions would signify the magnetically-induced binding of the 
excess electron by a neutral complex. 

\section*{Acknowledgments}

VGB is grateful for support from the Deutsche Forschungemeinschaft, as well 
as for support from the Universit{\'e} Paris-Sud during his visit at the 
Laboratoire Aim{\'e} Cotton in Orsay.
GGP acknowledges support by the National Aeronautics and Space 
Administration through Chandra Awards TM6-7005 and TM5-16005 issued by the Chandra X-ray 
Observatory Center, which is operated by the Smithsonian Astrophysical Observatory for 
and on behalf of the National Aeronautics Space Administration under contract NAS8-03060. 
The authors thank the anonymous referee for the valuable comments. 

\appendix

\section{The Landau states}

The quantum states for the Hamiltonian~(\ref{H1}) are the common eigenstates of 
the operators ${\mathit \Pi}_\perp^2 = {\mathit \Pi}_x^2 + {\mathit \Pi}_y^2$, 
$K_\perp^2 = K_x^2 + K_y^2$, and $L_z$, where 
\begin{align}
({\mathit \Pi}_x,{\mathit \Pi}_y,0) & = \vv{P}_\perp - (Q/2)\vv{B}\times\vv{R}_\perp~, 
\label{Pi}
\\
(K_x,K_y,0)                         & = \vv{P}_\perp + (Q/2)\vv{B}\times\vv{R}_\perp~, 
\label{K}
\\
(0,0,L_z)                           & = \vv{R}_\perp \times \vv{P}_\perp 
\label{L}
\end{align}
are the {\em kinetic}, {\em pseudo}, and {\em angular} momenta, respectively, 
for the motion of a particle with charge $Q$ in the plane perpendicular to the 
magnetic field. Their eigenvalues are determined by the integers:  
\begin{align}
{\mathit \Pi}_\perp^2 & = |Q|B(2N+1)~,  & N   &= 0,1,2,\ldots~, 
\label{Pi-vals} 
\\
K_\perp^2             & = |Q|B(2N_0+1)~,& N_0 &= 0,1,2,\ldots~, 
\label{K-vals}
\\
L_z                   & = -\sigma L~,   & L   &= 0,\pm 1,\pm 2,\ldots~,
\label{L-vals}
\end{align}
where $\sigma$ is the sign of $Q$. 

The operator ${\mathit \Pi}_\perp^2$ determines the kinetic energy of the particle 
given by the Hamiltonian~(\ref{H1}), whereas the operators $K_\perp^2$ and $L_z$ 
are the {\em integrals of motion} for this Hamiltonian. 
In particular, $K_\perp^2$ relates to the guiding center of the particle's rotation 
around the field lines, see, e.g., \cite{AHS_1981}. The eigenvalues~(\ref{Pi-vals}) 
yield the energies~(\ref{Lan-en}) for the Landau states, which are degenerate 
with respect to the numbers $N_0$ and $L$. 

Only two of the three numbers $N$, $N_0$ and $L$ are independent, and 
the relation $N=N_0+L$ (see Eq.~(\ref{relation-numbers})) holds, as a result 
of the operator relation 
\be 
{\mathit \Pi}_\perp^2 = K_\perp^2 - 2QBL_z~. 
\label{relation-operators}
\ee 
Thus, any pair from the set $N$, $N_0$ and $L$ can be used to designate the 
eigenstates of the Hamiltonian~(\ref{H1}). In our considerations, we use the 
pair $N$, $N_0$.

For a particle with charge $Q$ occupying a Landau state with the quantum numbers 
$N$ and $N_0$, the normalized wave function in ${\bf R}_\perp=(X,Y)$ is 
\be
{\mathit \Phi}^{(Q)}_{N,N_0}({\bf R}_\perp) = 
\frac{e^{{\rm i}{\sigma}(N_0-N)\varphi}}{\sqrt{2\pi}\lambda_Q}\,F_{N,N_0}(u),
\label{LF-cartesian}
\ee
where $\varphi=\tan^{-1}(Y/X)$, $u=(X^2+Y^2)/(2\lambda_Q^2)$, and 
$\lambda_Q=(|Q|B)^{-1/2}$. 
The functions $F_{N,N_0}$ are equal to zero if $N<0$ or $N_0<0$, while for 
non-negative indices they are given by the expression
\bea
F_{N,N_0}(u) &=& (-1)^{N-N_0} F_{N_0,N}(u)
\nonumber \\ 
             &=& \left(\frac{N!}{N_0!}u^{N_0-N}{\rm e}^{-u}\right)^{1/2}L_{N}^{N_0-N}(u),
\label{F-functions}
\eea
where $L_N^s(u)$ are the Laguerre polynomials. 
The set of Landau functions~(\ref{LF-cartesian}) with $N=0,1,2,\ldots$, 
$N_0=0,1,2,\ldots$ possesses the standard properties of orthogonality and completeness. 
The relations 
\begin{align}
({\mathit \Pi}_x-{\rm i}\,\sigma{\mathit \Pi}_y){\mathit \Phi}^{(Q)}_{N,N_0} &= 
-{\rm i}\,p_Q \sqrt{N+1}   \,{\mathit \Phi}^{(Q)}_{N+1,N_0},
\nonumber \\
({\mathit \Pi}_x+{\rm i}\,\sigma{\mathit \Pi}_y){\mathit \Phi}^{(Q)}_{N,N_0} &= 
\;\;\; {\rm i}\,p_Q \sqrt{N}     \,{\mathit \Phi}^{(Q)}_{N-1,N_0},
\nonumber \\
(K_x-{\rm i}\,\sigma K_y){\mathit \Phi}^{(Q)}_{N,N_0} &= 
-{\rm i}\,p_Q \sqrt{N_0}   \,{\mathit \Phi}^{(Q)}_{N,N_0-1},
\nonumber \\
(K_x+{\rm i}\,\sigma K_y){\mathit \Phi}^{(Q)}_{N,N_0} &= 
\;\;\; {\rm i}\,p_Q \sqrt{N_0+1} \,{\mathit \Phi}^{(Q)}_{N,N_0+1}, 
\label{cre/ann}
\end{align}
where $p_Q = (2|Q|B)^{1/2}$, are convenient for calculating matrix elements with 
the Landau functions.

The cyclotron transition amplitudes are determined by the matrix elements 
$D^{(\beta)}_{N',N}$ of the dipole operators~(\ref{dipop}) with the Landau 
functions~(\ref{LF-cartesian}). The corresponding transition energies are 
$\omega^{\rm cyc}_{N',N} = E^{\rm Lan}_{N'}(\Omega) - E^{\rm Lan}_N(\Omega)$, 
and the oscillator strengths are 
\be
f^{\rm cyc}_{N',N} = 2\,\omega^{\rm cyc}_{N',N}\left| D^{(\beta)}_{N',N} \right|^2~.
\label{f-cyc-def}
\ee
Direct calculations show that the non-vanishing matrix elements are 
\be
D^{(\beta)}_{N-\beta\sigma,N} = \langle {\mathit \Phi}^{(Q)}_{N-\beta\sigma,N_0} 
                                     \left| D^{(\beta)} \right| 
                                     {\mathit \Phi}^{(Q)}_{N,N_0} \rangle~,  
\label{D-cyc}
\ee
and therefore the selection rules and transition energies are given 
by Eqs.~(\ref{selection-cyc}) and (\ref{omega-cyc}), respectively. 
The matrix elements~(\ref{D-cyc}) are readily evaluated 
using the relations~(\ref{cre/ann}). Four cases, 
selected by the signs of the charge $\sigma=\pm 1$ and circular polarization 
$\beta = \pm 1$, cast into the expression~(\ref{f-cyc}). 

The matrix elements~(\ref{D-cyc}) are calculated with pairs of 
the Landau functions with the same quantum number $N_0$. Since this 
number does not change in the course of the cyclotron transitions, we 
do not include it in the indices for the dipole matrix elements. 
Notice that the dipole matrix elements also differ from zero when being 
calculated with the Landau functions with neighboring values for $N_0$ and the 
same values for $N$. They relate then to the zero transition energies 
and do not describe physically meaningful transitions. 

\section{System of charges in a magnetic field: integrals of motion and selection rules}

Similar to the one-particle Hamiltonian~(\ref{H1}), the integrals of motion for the 
many-particle Hamiltonian~(\ref{H-many}) are determined by the 
operators $K_\perp^2 = K_x^2 + K_y^2$ and ${\cal L}_z$ that are now the {\em total} 
pseudo- and angular momenta for the motion of the system of particles in the plane 
perpendicular to the magnetic field, 
\begin{align}
(K_x,K_y,0)      & = 
\sum_a \left( \vv{p}_{\perp{a}} + \frac{e_a}{2}\,\vv{B}\times\vv{r}_{\perp{a}} \right)~, 
\label{K-tot}
\\
(0,0,{\cal L}_z) & = \sum_a {\bf r}_{\perp{a}} \times {\bf p}_{\perp{a}}~.
\label{L-tot}
\end{align}
The many-particle pseudo-momentum~(\ref{K-tot}) can be transformed to the 
{\em one-particle form}~(\ref{K}) when expressed in terms of an appropriate 
canonical pair ${\bf R}_\perp$, ${\bf P}_\perp$. This pair can be introduced in 
several ways. For example, it can be related to the center of charge for the system. 
Another choice is to relate ${\bf R}_\perp$, ${\bf P}_\perp$ to the c.m.\ motion 
followed by a gauge transformation of the operators. 

As the total pseudo-momentum can be transformed to the one-particle form, we 
keep the same notation for this operator as introduced for the single charge. 
We also keep the notation $N_0$ for the quantum number which determines the 
eigenvalues of $K_\perp^2$. The latter eigenvalues are given by 
Eq.~(\ref{K-vals}) applicable to many-particle systems. 
The eigenvalues of ${\cal L}_z$ are integers which we specify by the relation 
(cf. Eq.~(\ref{L-vals}))   
\begin{align}
\Lt_z& = -\sigma\Lt~,& \Lt& = 0,\pm 1,\pm 2,...~.
\label{L-tot-vals}
\end{align}

The dipole transitions for the system are determined by the operators~(\ref{dip}). 
To derive the selection rules, we consider the matrix elements $\bm{\D}_{q'q}$ 
for the dipole moment~(\ref{DipMom}), where $q$ and $q'$ denote the quantum 
states with the numbers $N_0$, $\Lt$ and $N'_0$, $\Lt'$, respectively. 

The eigenvalues~(\ref{K-vals}) and (\ref{L-tot-vals}) allow one to 
relate $\bm{\D}_{q'q}$ to the matrix elements of the commutators,
\bea
p_Q^2 \left( N'_0 - N_0 \right) \bm{\D}_{q'q} &=& \left[ K_\perp^2, \bm{\D} \right]_{q'q}~,
\label{r1} 
\\
-\sigma \left( \Lt' - \Lt \right) \bm{\D}_{q'q} &=& \left[ \Lt_z, \bm{\D} \right]_{q'q}~.
\label{r2}
\eea
By evaluating the commutators of the Hamiltonian~(\ref{H-many}) with the radius 
$\vv{r}_a$, one obtains the relation 
\be
\left( E_{q'} - E_q \right) \bm{\D}_{q'q} = {\rm i}\,\bm{\mathcal{J}}_{q'q}~,
\label{relation}
\ee
where $E_{q'}$ and $E_q$ are the state energies and $\bm{\mathcal{J}}_{q'q}$ are 
the matrix elements for the current 
\be
\bm{\mathcal{J}} = \sum_a \frac{e_a}{m_a} 
                   \left( \vv{p}_a - \frac{e_a}{2}\,\vv{B}\times\vv{r}_a \right)~. 
\label{current}
\ee

To derive the selection rules for the numbers $N'_0$ and $N_0$, we 
use Eq.~(\ref{relation}) to transform the relation~(\ref{r1}) as follows 
\be 
p_Q^2 \left( N'_0 - N_0 \right) \left( E_{q'} - E_q \right) \bm{\D}_{q'q} = 
{\rm i}\,\left[ K_\perp^2, \bm{\mathcal{J}} \right]_{q'q}~.
\label{r1a} 
\ee
It is straightforward to prove that the current commutes with $K_\perp^2$ and therefore 
\be 
\left( N'_0 - N_0 \right) \left( E_{q'} - E_q \right) \bm{\D}_{q'q} = 0~.
\label{r1b} 
\ee
For the physically meaningful transitions, the state energies are different, 
$E_{q'} \ne E_q$. Therefore, the transitions with $\bm{\D}_{q'q} \neq 0$ 
correspond to the selection rule 
\be
N'_0 = N_0~.
\label{selection-N_0}
\ee
Notice that this general selection rule holds for an arbitrary polarization of 
radiation, i.e. for the transitions determined not only by the cyclic components~(\ref{dip}) 
but by projection of the dipole operator on an arbitrary polarization vector. 
On physical grounds, the number $N_0$ relates to the ``center of quantization'' in uniform 
space. Therefore, a change in $N_0$ implies the change of the reference frame, not 
affecting the properties of the quantum states. 

To derive the selection rules with respect to the numbers $\Lt'$ and $\Lt$, 
we consider the relations~(\ref{r2}) for the cyclic components of the dipole 
operator, 
\be
\left( \Lt' - \Lt \right) \D^{(\beta)}_{q'q} = 
-\sigma \left[ \Lt_z, \D^{(\beta)} \right]_{q'q}~,
\label{r2a}
\ee
and use the commutator relations 
\be 
[\Lt_z,\D^{(\beta)}] = \beta\,\D^{(\beta)} 
\label{commutator}
\ee 
which are easily proved by direct calculations. 
Then, Eqs.~(\ref{r2a}) and (\ref{commutator}) yield the relation 
\be
\left( \Lt' - \Lt + \beta\sigma \right) \D_{q'q}^{(\beta)} = 0~. 
\label{r2b}
\ee
Therefore, the selection rules for the transitions with 
$\D_{q'q}^{(\beta)} \neq 0$ are 
\be
\Lt' = \Lt - \beta\sigma~. 
\label{selection-L}
\ee
As the quantum number $N_0$ does not change in the course of transitions, 
the same selection rule, $\N'=\N-\beta\sigma$, Eq.~(\ref{selection-cyc-a}), 
holds for the numbers ${\cal N}'$ and ${\cal N}$.

\section{Perturbation treatment of the coupling between c.m.\ and electronic motions} 

When employing standard perturbation techniques to treat the coupling term $H_3$ in 
the Hamiltonian~(\ref{H}), we implement the integrals of motion 
$K_\perp^2$ and $\Lt_z$ already for the zero-order approximation 
(which is possible since the operators $K_\perp^2$ and $\Lt_z$ commute 
with the terms $H_1$ and $H_2$). The perturbation-corrected ion states emerge then 
as the eigenstates of the integrals of motion as well. 

The zero-order approximation to the Hamiltonian~(\ref{H}) is the sum $H_1+H_2$, 
the corresponding wave functions are the products 
\be
\psi^{(0)}_{NN_0s\nu}\left(\vv{R}_\perp,\left\{\vv{r}_i\right\}\right) = 
{\mathit \Phi}^{(Q)}_{N,N_0}\left(\vv{R}_\perp\right)
 \phi_{s\nu}\left(\left\{\vv{r}_i\right\}\right)
\label{wfs-zero}
\ee
of the eigenfunctions of the terms $H_1$ and $H_2$, and the ion energies are 
the sums~(\ref{ens-zero}) of the c.m.\ and electronic energies. As the canonically 
transformed total pseudo-momentum is the one-particle operator~(\ref{K}) for the 
c.m.\ motion, the c.m.\ Landau functions ${\mathit \Phi}^{(Q)}_{N,N_0}$ and hence 
the zero-order wave functions~(\ref{wfs-zero}) are the eigenfunctions of 
$K_\perp^2$. The corresponding quantum number $N_0$ is explicitly included in the 
set of the numbers that label the zero-order states. Since the electronic wave 
functions $\phi_{s\nu}(\{{\bf r}_i\})$ are the eigenfunctions of $l_z$ with the 
eigenvalues $-s$, the zero-order wave functions are the eigenfunctions of $\Lt_z$ 
with the eigenvalues $-\sigma(N-N_0) - s$ (cf. Eqs.~(\ref{L-vals}), 
(\ref{relation-numbers}) and (\ref{Lz-ion})). Thus, to ascribe the conserved values 
$-\sigma\Lt$ (see Eq.~(\ref{L-tot-vals})) of the total longitudinal angular momentum 
to the zero-order states, the relation $N = N_0 + \Lt - \sigma{s}$ is required, which 
yields $N = \N -\sigma{s}$ (cf. Eq.~(\ref{N-ion})). Notice that the zero-order wave 
functions depend on the sum $\N=N_0+\Lt$ (cf. Eq.~(\ref{relation-numbers-1})) 
providing thereby the same property for the perturbation-corrected wave functions. 

The perturbation corrections are determined by the matrix elements 
\begin{align}
& \langle\psi^{(0)}_{\N{N_0}{s'}\nu'}|H_3|\psi^{(0)}_{\N{N_0}{s}\nu}\rangle = 
-\frac{\alpha|Q|^{1/2}B^{3/2}}{M}\,
\label{H3-matrix} \\
&\times \left[ \sqrt{N+1}\,d^{(+\sigma)}_{s'\nu',s\nu}\,\delta_{s',s-\sigma}
             + \sqrt{N}\,  d^{(-\sigma)}_{s'\nu',s\nu}\,\delta_{s',s+\sigma} \right]
\nonumber
\end{align}
obtained with use of the properties~(\ref{cre/ann}) for the c.m.\ Landau functions, with 
$d^{(\pm\sigma)}_{s'\nu',s\nu} = \langle\phi_{s'\nu'}|d^{(\pm\sigma)}|\phi_{s\nu}\rangle$ 
being the matrix elements of the cyclic components of electronic dipole operator 
($\sigma$ is the sign of the ion's charge). From the 
commutator relations $[l_z,d^{(\pm 1)}]=\pm d^{(\pm 1)}$ it follows that 
non-vanishing are the elements $d^{(\pm 1)}_{s\mp{1},\nu';s\nu}$ as reflected by 
the Kronecker deltas in Eq.~(\ref{H3-matrix}). 

Since the perturbation matrix elements with $s'=s$ are equal to zero, the 
first-order energy corrections vanish. Standard calculations of the second-order 
energy corrections yield 
\bea
E^{(2)}_{\N{s}\nu} &=& \frac{\alpha^2|Q|B^3}{M^2} 
                       \left[ (N+1) G_{s\nu}^{(+)} + N G_{s\nu}^{(-)} \right], 
\nonumber 
\\
G_{s\nu}^{(+)} &=& \sum_{\nu'\neq\nu} 
                   \frac{\left|d^{(+\sigma)}_{s-\sigma,\nu';s\nu}\right|^2}
                        {\E_{s\nu}-\E_{s-\sigma,\nu'}-\Omega}~,
\nonumber 
\\
G_{s\nu}^{(-)} &=& \sum_{\nu'\neq\nu} 
                   \frac{\left|d^{(-\sigma)}_{s+\sigma,\nu';s\nu}\right|^2}
                        {\E_{s\nu}-\E_{s+\sigma,\nu'}+\Omega}~. 
\label{ens-second} 
\eea
Similar to the zero-order energies~(\ref{ens-zero}), the corrections 
$E^{(2)}_{\N{s}\nu}$ depend linearly on $N=\N-\sigma{s}$. 
Therefore, the sum of the zero- and second-order terms results in the 
oscillator-like dependence~(\ref{ens-result}) of the ion energies on $N$. 
As a result of the coupling between c.m.\ and internal motions, 
the ion mass $M$ in the Landau energies is replaced by the effective mass $M_{s\nu}$ 
dependent on the internal states (cf. Eqs.~(\ref{Omega_eff})). 
In addition, the ion energies acquire the shifts $\Delta_{s\nu}$ 
with respect to the zero-order energies~(\ref{ens-zero}). The effective 
masses and shifts calculated from the second-order perturbation correction are
\bea
\frac{M}{M_{s\nu}} &=& 1 + \frac{\alpha^2B^2}{M} \left[ G_{s\nu}^{(+)} + G_{s\nu}^{(-)} \right],
\label{mass-ratio} 
\\
\Delta_{s\nu}      &=& \frac{\alpha^2|Q|B^3}{M^2}\,G_{s\nu}^{(+)}~. 
\label{shift}
\eea

It now remains to account for the perturbation corrections for the oscillator 
strengths. For transparency of the derivations we consider the bound-ion 
emission cyclotron transitions with $\N' = \N - 1$ for the positive ion, 
$\sigma=+1$. The emitted radiation is right-polarized, $\beta=+1$. With accuracy 
up to the first-order perturbation corrections, the initial and final 
state wave-functions are 
\bea
\psi_{\N{N_0}{s}\nu} &=& {\mathit \Phi}^{(Q)}_{N,N_0}\,\phi_{s\nu}  
       -  \sqrt{N+1}\,{\mathit \Phi}^{(Q)}_{N+1,N_0}\tilde{\phi}_{s-1,\nu} 
\nonumber \\
      &-& \sqrt{N}\,  {\mathit \Phi}^{(Q)}_{N-1,N_0}\tilde{\phi}_{s+1,\nu}~,
\nonumber \\
\psi_{\N'{N_0}{s}\nu} &=& {\mathit \Phi}^{(Q)}_{N-1,N_0}\,\phi_{s\nu}  
       -  \sqrt{N}\,  {\mathit \Phi}^{(Q)}_{N,N_0}  \tilde{\phi}_{s-1,\nu} 
\nonumber \\
      &-& \sqrt{N-1}\,{\mathit \Phi}^{(Q)}_{N-2,N_0}\tilde{\phi}_{s+1,\nu}~,
\label{if}
\eea
where $N=\N-s$ (cf. Eq.~(\ref{N-ion})), and the functions 
\be
\tilde{\phi}_{s\pm{1},\nu} = \frac{\alpha|Q|^{1/2}B^{3/2}}{M}\,
                             \sum_{\nu'\neq\nu} 
                             \frac{d^{(\mp{1})}_{s\pm{1},\nu';s\nu}\,\phi_{s\pm{1},\nu'}}
                                  {\E_{s\nu} - \E_{s\pm{1},\nu'} \pm \Omega}
\label{tildes}
\ee
include the electronic configurations admixed to the zero-order states 
as a result of the coupling. Due to the 
latter admixtures, the matrix elements of the dipole operators 
$\D^{(+1)}=D^{(+1)}+d^{(+1)}$ are contributed by the electronic transitions,
\be 
d^{(+1)}_{\N',\N} = -\frac{\alpha|Q|^{1/2}B^{3/2}}{M}\,\sqrt{N}\,
                     \left[G_{s\nu}^{(+)} + G_{s\nu}^{(-)} \right], 
\label{contribution}
\ee
in addition to the dominant c.m.\ cyclotron transitions, 
\be
D^{(+1)}_{\N',\N} = -\sqrt{|Q|/B}\,\sqrt{N}~.
\label{cmcyc}
\ee 
The sums of the contributions~(\ref{cmcyc}) and (\ref{contribution}) yield the 
total dipole elements different from the c.m.\ ones~(\ref{cmcyc}) by exactly the 
ratio of masses given by Eq.~(\ref{mass-ratio}):  
\be
\D^{(+1)}_{\N',\N} = (M/M_{s\nu})\,D^{(+1)}_{\N',\N}~.
\label{D-results}
\ee
With account for the mass ratio factor for the transition energies 
$\omega_{\N'\!,\,\N} = (M/M_{s\nu})\,\omega^{\rm cyc}_{N'\!,N}$, we arrive at 
the relation $f_{\N'\!,\,\N} = (M/M_{s\nu})^3 f^{\rm cyc}_{N'\!,N}$ for the 
oscillator strengths (see Eq.~(\ref{cyc-second})). 

\section{Coupled-channel formalism}

The equations below apply to the positive hydrogen-like ions and negative 
magnetically-induced ions studied numerically in Section~V. 

\subsection{Hydrogen-like ions}

As the basis states for the Hamiltonian~(\ref{H_cc}), we employ the Landau states 
of separated electron and nucleus, i.e., the states of a detached ion. The states 
are attributed to the quantum numbers $N_0$ and $\Lt$ for the integrals of collective 
motion~\cite{B_1995} and describe the electron and nucleus occupying the Landau levels 
with the numbers $n_e$ and $n$, respectively. The corresponding energies are the sums 
$E^{\rm Lan}_{n_e}(\Omega_e) + E^{\rm Lan}_n(\Omega_0)$, where 
$\Omega_e = B$ and $\Omega_0 = ZB/M_0$ are the electron and nucleus cyclotron energies, 
respectively ($Z$ is the nucleus charge number and $M_0$ is the nucleus mass). 
For given $\N=N_0+\Lt$, the Landau level numbers vary as 
\be
n_e = 0,1,\ldots~,\;\;\;
n = 0,1,\ldots,\N+n_e~, 
\label{ne_n_variation} 
\ee 
and the basis functions with the numbers $n_e$ and $n$ are given 
by the linear combinations 

\begin{align}
& \langle {\bf R}_\perp,{\bf r}_\perp|n_e n \rangle = 
\nonumber \\
& \sum_{k=-n_e}^\N c_{n,n_e+k}^{\N+n_e} 
{\mathit \Phi}^{(Z-1)}_{\N-k,N_0}({\bf R}_\perp)\, 
{\mathit \Phi}^{(-1)}_{n_e,n_e+k}({\bf r}_\perp)
\label{n-chnls}
\end{align}
of the Landau functions for the particles with charges $Z-1$ of the entire ion 
and $-1$ of the electron (in units of the elementary charge $|e|$). 

The coefficients $c_{nk}^j$ with $n=0,1,\ldots,j$, $k=0,1,\ldots,j$ 
were introduced in Ref.~\cite{B_1995} as generated from the recursion and 
normalization relations 
\begin{align}
& [(Z-1)j - Zn -(Z-2)k]\,c_{nk}^j = \sqrt{Z-1} 
\nonumber \\
& \times \left[ \sqrt{(j-k+1) k}\,c_{n,k-1}^j + \sqrt{(j-k)(k+1)}\,c_{n,k+1}^j \right]~,
\nonumber \\
& \sum_{k=0}^j \left[ c_{nk}^j \right]^2 = \sum_{n=0}^j \left[ c_{nk}^j \right]^2 = 1~,
\label{rec_norm}
\end{align}
and shown to possess the orthogonality and completeness properties
\be
\sum_{k=0}^j c_{nk}^j c_{n'k}^j = \delta_{nn'}~,
\;\;\;
\sum_{n=0}^j c_{nk}^j c_{nk'}^j = \delta_{kk'}~.
\label{properties}
\ee
A further analysis allows to derive additional useful relations for $c_{kn}^j$: 
\bea
\sqrt{n+1}\,c_{n+1,k}^{j+1} &=& a\,\sqrt{j+1-k}\,c_{nk}^j - b\,\sqrt{k}\,  c_{n,k-1}^j~,
\nonumber \\ 
\sqrt{n}\,c_{n-1,k}^{j-1}   &=& a\,\sqrt{j-k}\,  c_{nk}^j - b\,\sqrt{k+1}\,c_{n,k+1}^j~,
\nonumber \\ 
\sqrt{j-n+1}\,c_{nk}^{j+1}  &=& b\,\sqrt{j+1-k}\,c_{nk}^j + a\,\sqrt{k}\,  c_{n,k-1}^j~,
\nonumber \\ 
\sqrt{j-n}\,c_{nk}^{j-1}    &=& b\,\sqrt{j-k}\,  c_{nk}^j + a\,\sqrt{k+1}\,c_{n,k+1}^j~,
\nonumber \\
a = \sqrt{Z-1}/\sqrt{Z}     &,& \;\; b = 1/\sqrt{Z}~, \;\; a^2+b^2=1~.
\label{rec_add}
\eea

The wave function for the ion with given $N_0$ and $\Lt$ is expanded in the basis 
set~(\ref{n-chnls}) as 
\be
  \psi({\bf R}_\perp,{\bf r}_\perp,z) 
= \sum_{n_e=0}^\infty \sum_{n=0}^{\N+n_e} 
  \langle {\bf R}_\perp,{\bf r}_\perp|n_e n \rangle \, g_{n_e n}(z)~.
\label{nen-expansion}
\ee
For calculations of the states and cyclotron transitions of He$^+$ at the magnetic 
field strengths typical for the atmospheres of neutron stars, we employ the 
adiabatic approximation by restricting the basis to the states with $n_e=0$. We count 
the ion energy from the detachment threshold 
$E^{\rm Lan}_0(\Omega_e) + E^{\rm Lan}_0(\Omega_0) = \Omega_e/2 + \Omega_0/2$, 
and the adiabatic approximation is justified by the condition $E \ll \Omega_e=B$ for 
the energies computed. The coupled-channel equations for the energies and 
functions $g_n(z)=g_{0n}(z)$, $n=0,1,\ldots,\N$, are 
\begin{align}
&-\frac{1}{2\mu}\,\frac{{\rm d}^2 g_n(z)}{{\rm d} z^2} 
 + \Big[ V_{nn}(z) + n\Omega_0 - E \Big] g_n(z)
 \nonumber \\
&= -\mathop{\sum_{n'=0}^\N}_{n' \neq n} 
    V_{nn'}(z)\,g_{n'}(z)~,
\label{n-eqs}
\end{align}
where $\mu = M_0/(1+M_0)$ is the reduced mass ($M_0 \gg 1$ and $\mu \approx 1$ in 
the atomic units of mass). 

The right-hand side of Eq.~(\ref{n-eqs}) involves the potentials $V_{nn'}(z)$ which 
are the projections of the electron-nucleus Coulomb interaction $-Z/r$ onto the basis 
states~(\ref{n-chnls}) with $n_e=0$, 
\be
V_{nn'}(z) = -Z \sum_{k=0}^\N \frac{c_{n'k}^\N c_{nk}^\N}{k!}
                \int_0^\infty \frac{u^k\exp(-u){\rm d}u}{\sqrt{(2u/B)+z^2}}~.
\label{nn'-potentials}
\ee
Since the potentials depend on $\N=N_0+\Lt$ but not on the numbers $N_0$ and $L$ 
separately, the same property holds for the energies and the channel functions 
found from solving Eqs.~(\ref{n-eqs}). This is a general property of the 
coupled-channel formalism not restricted by the adiabatic approximation employed.   
The multi-channel wave functions~(\ref{nen-expansion}) depend on both $\N$ and $N_0$, 
as the basis functions~(\ref{n-chnls}) depend on both numbers. The quantum states 
of the moving ion are degenerate with respect to $N_0$ reflecting the fact that 
this number relates to locations of the guiding center for the ion in a homogeneous 
space. 

The numerical integration of the coupled-channel equations~(\ref{n-eqs}) is performed 
to match the boundary conditions at $z=0$ and $|z|\to\infty$ which are determined by 
the parity and the asymptotic decay of the channel functions, respectively, and given 
by the equations below. As the 
potentials~(\ref{nn'-potentials}) are even functions of $z$, the ion states can be 
attributed to a given $z$-parity which is even for the tightly-bound states: 
\be
{\rm d}g_n(z)/{\rm d}z = 0 \;\; \mbox{for} \;\; z=0~. 
\label{bc1}
\ee
With increasing $|z|$, the off-diagonal potentials decrease faster than the diagonal 
ones which converge to a one-dimensional Coulomb potential $-Z/|z|$. The 
equations~(\ref{n-eqs}) decouple, and the decay of the bound channel functions 
follows the relation 
\be
\frac{{\rm d}g_n(z)/{\rm d}z}{g_n(z)} 
= -\kappa_n\,\frac{2z - 1 - \kappa_n}{2z + 1 - \kappa_n} 
\label{ld}
\ee
at $|z| \gg {\rm max}(1,\kappa_n)$ (the unit of distance is the Bohr radius), 
where $\kappa_n = \sqrt{2(n\Omega_0-E)}$.
 
The solutions of the equations~(\ref{n-eqs}) are the ion energies $E_{\N{s}}$ and the 
multi-channel wave functions 
\be
\psi_{\N{s}} = \sum_{n=0}^\N \langle {\bf R}_\perp,{\bf r}_\perp|0 n \rangle \, g_{\N{sn}}(z)~,
\label{n-expansion}
\ee
where $s=0,1,\ldots,\N$ enumerates the solutions for a given $\N$. We can assume 
that for different $\N$ the solutions with the same $s$ correspond to the similar internal 
excitations, see the discussion before Eq.~(\ref{correspondence}) in Sec.~IV. The bound-ion 
cyclotron transitions are therefore the transitions between the states represented by the 
sets of channel functions $\{g_{\N{sn}}(z)\}_{n=0}^\N$ and 
$\{g_{\N\pm{1},{sn'}}(z)\}_{n'=0}^{\N\pm{1}}$. By using the properties~(\ref{cre/ann}) of the 
Landau functions, the dipole matrix elements for these transitions are calculated as follows 
\begin{align}
& {\mathcal D}^{(+1)}_{\N-1,\N} = 
-D_0 \sum_{n'=0}^{\N-1}\sum_{n=0}^\N \langle g_{\N-1,n'}|g_{\N{n}} \rangle \sum_{k=0}^\N c_{nk}^\N 
\nonumber \\ 
& \times \left[ a\,\sqrt{\N-k}\,c_{n'k}^{\N-1} + b\,\sqrt{k}\,c_{n',k-1}^{\N-1} \right]~,
\label{step1-em}
\\
& {\mathcal D}^{(-1)}_{\N+1,\N} = 
-D_0 \sum_{n'=0}^{\N+1}\sum_{n=0}^\N \langle g_{\N+1,n'}|g_{\N{n}} \rangle \sum_{k=0}^\N c_{nk}^\N 
\nonumber \\ 
& \times \left[ a\,\sqrt{\N-k+1}\,c_{n'k}^{\N+1} + b\,\sqrt{k+1}\,c_{n',k+1}^{\N+1} \right]~,
\label{step1-abs}
\end{align}
where $D_0 = \sqrt{Z/B}$ and the coefficients $a$ and $b$ are given by Eqs.~(\ref{rec_add}). 
The numbers $s$ are omitted from the notations of the channel functions to simplify the 
equations. The recurrence relations~(\ref{rec_add}) allow one to replace the expressions 
in square brackets in Eqs.~(\ref{step1-em}) and (\ref{step1-abs}) by $\sqrt{n'+1}\,c_{n'+1,k}^\N$ 
and $\sqrt{n'}\,c_{n'-1,k}^\N$, respectively. The summations over $k$ and $n'$ can then be 
performed making use of the orthogonality properties~(\ref{properties}) of the 
$c$-coefficients. This results in the formulas 
\begin{align}
& {\mathcal D}^{(+1)}_{\N-1,\N} = -D_0 \sum_{n=0}^\N \sqrt{n}\,\langle g_{\N-1,n-1}|g_{\N{n}} \rangle~,
\label{D-em-result}
\\
& {\mathcal D}^{(-1)}_{\N+1,\N} = -D_0 \sum_{n=0}^\N \sqrt{n+1}\,\langle g_{\N+1,n+1}|g_{\N{n}} \rangle
\label{D-abs-result}
\end{align}
convenient for the multi-channel calculations of the bound-ion cyclotron transitions. 

\subsection{Magnetically-induced anions}

In contrast to positive ions, the states of a detached negative ion are not discrete because 
the motion of the separated neutral system is not confined by the magnetic field. Although such states 
can also be constructed as the eigenstates of $K_\perp^2$ and ${\cal L}_z$~\cite{BC2003}, expanding 
the ion wave function in this basis involves integration over the continuum related to the motion of 
the neutral core. This yields the coupled-channel equations coupled by an integral operator which 
complicates the numerical treatments. As a workaround, we use a discrete basis set~\cite{BSC2007} 
describing the electron occupying the Landau level $n_e$ with the longitudinal angular 
momentum $l_z = -j$, $j=0,\pm{1},\pm{2},\ldots$. We designates the values of $l_z$ by 
the quantum number $j$ instead of $s$ used in Eq.~(\ref{l-vals}), because $s$ is now used to 
enumerate the multi-channel quantum states. The basis set is also attributed to the 
quantum numbers $N_0$ and $\Lt$ and therefore depends on the transverse coordinates of both 
electron and neutral system, 
\be
\langle {\bf R}_\perp,{\bf r}_\perp|n_e j \rangle = 
{\mathit \Phi}^{(-1)}_{\N+j,N_0}({\bf R}_\perp)\, 
{\mathit \Phi}^{(-1)}_{n_e,n_e+j}({\bf r}_\perp)~.
\label{j-chnls}
\ee
For a given $\N=N_0+\Lt$, the ion wave function is expanded in the basis as 
\be
  \psi({\bf R}_\perp,{\bf r}_\perp,z) 
= \sum_{n_e=0}^\infty \sum_{j=j_{\rm min}}^\infty 
  \langle {\bf R}_\perp,{\bf r}_\perp|n_e j \rangle \, g_{n_e j}(z)~,
\label{nej-expansion}
\ee
where $j_{\rm min} = \max(-\N,-n_e)$ (the lower boundary for $j$ follows from the 
restrictions $\N+j \ge 0$ and $n_e+j \ge 0$). The ion energy is counted from the detachment 
threshold $E^{\rm Lan}_0(\Omega_e) = \Omega_e/2$. 

Similar to hydrogen-like ions in strong magnetic fields, we employ the adiabatic 
approximation by including only the channels with $n_e=0$ into the wave function 
expansion~(\ref{nej-expansion}). This approximation is justified for magnetically-bound 
anions at laboratory magnetic field strengths by smallness of the energies compared to the 
electron Landau energy. The coupled-channel equations for the ion energy $E$ and the functions 
$g_j(z)=g_{0j}(z)$, $j = j_{\rm min}, j_{\rm min}+1, \ldots$, where $j_{\rm min}=\max(-\N,0)$, 
are
\begin{align}
-\frac{1}{2\mu}\,\frac{{\rm d}^2 g_j(z)}{{\rm d} z^2} 
&+ \big[ V_j(z) + \Omega_0(\N+2j+1) - E \big] g_j(z)
\nonumber \\
&+ \Omega_0\sqrt{(\N+j+1)(j+1)}\,g_{j+1}(z)
\nonumber \\
&+ \Omega_0\sqrt{(\N+j)j}\,g_{j-1}(z) = 0~,
\label{j-eqs}
\end{align}
where $\Omega_0 = B/M_0$ is the cyclotron energy of a particle with the mass $M_0$ of 
the neutral core and the charge of the entire anion. 
The potential $V_j(z)$ is an average of the polarization potential $-\kappa/(2r^4)$ with 
the probability density determined by the basis functions~(\ref{j-chnls}) with $n_e=0$. 
It can presented in the form (see, e.g., Ref.~\cite{BSC2002})
\be
V_j(z) = - \frac{\kappa B^2}{8}\,
           \int_0^\infty {\rm d}\xi\,\frac{\xi\exp(-v\xi)}{(1+\xi)^{j+1}}~,
\label{V_j}
\ee
where $v = 0.5 B \left( z^2 + a^2 \right)$. The parameter $a$, which has a meaning of 
the size of neutral core, is introduced for a model cutoff of the polarization potential 
at $r \to 0$. 

We remark that Eqs.~(\ref{j-eqs}) depend on $\N=N_0+\Lt$, but not on the numbers 
$N_0$ and $\Lt$ separately, reflecting thereby a general property of the quantum states to 
be degenerate with respect to the number $N_0$. 

According to Eqs.~(\ref{j-eqs}), each channel couples to the neighboring ones 
due to the motion of the neutral system across the magnetic field, and the ``strength'' 
of the couplings is the cyclotron energy $\Omega_0$. The couplings remain the same 
at any $z$, and the channels do not decouple at infinite $|z|$ making it difficult to 
determine the asymptotic properties for the channel functions. This complicates direct 
numerical integration of the coupled-channel equations~(\ref{j-eqs}). To proceed with 
numerical treatments, we employ an expansion of the channel functions, 
\be
g_j(z) = \sum_{\nu=0}^\infty a_{j\nu}\varphi^{(\alpha)}_\nu(z),
\label{nu-expansion}
\ee
in the complete set of functions 
\be
\varphi^{(\alpha)}_\nu(z) = \sqrt{\alpha} \, 
                            \exp(-\alpha |z|) \, L_\nu(2 \alpha |z|),
\label{long}
\ee
where $\nu = 0,1,2,\ldots$, $L_\nu$ are the Laguerre polynomials and $\alpha$ is a 
parameter that determines spatial extensions of the functions. In particular, for 
$\alpha=(-2\E_s)^{1/2}$ where $\E_s$ is given by Eqs.~(\ref{binding_0})-(\ref{binding_s}), 
the function with $\nu=0$ describes the bound motion of the excess electron 
in the infinitely heavy anion (see Ref.~\cite{BSC2007} for the details). 

With the expansion~(\ref{nu-expansion}), the coupled-channel equations~(\ref{j-chnls}) can 
be transformed to linear equations for the expansion coefficients $a_{j\nu}$,
\be
\sum_{j=j_{\rm min}}^\infty \sum_{\nu=0}^\infty 
\left[ h_{j'\nu',j\nu} - E\,\delta_{j'j}\,\delta_{\nu'\nu} \right] a_{j\nu} = 0,
\label{leqs}
\ee
where
\bea
h_{j'\nu',j\nu} &=& \Omega_0\,T_{j'j}\,\delta_{\nu'\nu} 
                   +\left[ \frac{\alpha^2}{2\mu}\,t_{\nu'\nu} 
                           + V^{(j)}_{\nu'\nu} \right] \delta_{j'j}~, 
\nonumber \\ 
T_{j'j} &=& (\N+2j+1)\,\delta_{j'j} + \sqrt{(\N+j)j}\,\delta_{j',j-1}
\nonumber \\ 
        &+& \sqrt{(\N+j+1)(j+1)}\,\delta_{j',j+1}~,
\nonumber \\
t_{\nu'\nu} &=&  \delta_{\nu'\nu} + 2 (1 - \delta_{\nu'\nu}) + 4 \min\{\nu',\nu\}~,
\nonumber \\
V_{\nu'\nu}^{(j)} &=& \int_{-\infty}^\infty V_j(z)\, 
                      \varphi_{\nu'}^{(\alpha)}(z)\,\varphi_\nu^{(\alpha)}(z)\,{\rm d}z~.
\label{ingredients}
\eea
The next step is solving the equations~(\ref{leqs}), i.e. finding 
the eigenvalues and eigenvectors of a real symmetric matrix with the elements 
$h_{j'\nu',j\nu}$. In calculations, we restrict this matrix to the elements 
with $j$ and $j'$ varying from $j_{\rm min}$ to some maximal number $j_{\rm max}$, 
and $\nu$ and $\nu'$ varying from $0$ to some maximal number $\nu_{\rm max}$. 
The maximal numbers are increased until convergence of an eigenvalue $E$ of interest 
is achieved. For a more effective convergence the parameter $\alpha$ is varied  
to achieve a minimum value of $E$ 
for each set of numbers $j_{\rm max}$ and $\nu_{\rm max}$, which are varied 
in the course of iterative procedure. We remark that the functions~(\ref{long}) 
form a complete set at any $\alpha$ value, and that the computations converge to 
the energies independent of $\alpha$. 

As described in the previous parts of the paper, we label the ion energies by 
the number $s$ designating the internal excitations influenced by the coupling between 
the collective and internal motions. For a given $\N$, solving Eqs.~(\ref{leqs}) 
yields a {\em finite} number of the negative eigenvalues $E_{\N{s}}$ that form the 
$s$-branches of {\em bound} levels. Typically, we find a few branches for 
the atomic and cluster magnetically-induced anions (a maximal set of the branches 
for the anions discussed in Section~V is the set of $s=1,2,3,4,5$ branches of bound 
states for the Xe$_{13}^-$ anion at $B=50$~T). Along with the eigenvalues, we obtain the 
eigenvector components $a^{(\N{s})}_{j\nu}$ for the Hamiltonian matrix~$h_{j'\nu',j\nu}$, 
and can compute the multi-channel wave functions as 
\be
\psi_{\N{s}} = \sum_{j,\nu} \langle {\bf R}_\perp,{\bf r}_\perp|0 j \rangle 
                            a^{(\N{s})}_{j\nu} \varphi^{(\alpha)}_\nu(z)~,
\label{j-expansion}
\ee
where the limits of summations are determined as discussed above. 

According to the analysis presented in our studies, the bound-ion cyclotron transitions 
are the transitions between the states $\psi=\psi_{\N{s}}$ and $\psi'=\psi_{\N\pm{1},s}$. 
When the pair of states is computed with the same values of the parameter $\alpha$ 
(as it is done in our calculations), the transition matrix elements can be evaluated 
according to the expressions 
\bea
\sqrt{B}\,D^{(+1)}_{\N+1,\N} &=& \sum_{j,\nu} \sqrt{j}\,a^{(\N+1,s)}_{j-1,\nu} a^{(\N{s})}_{j\nu} 
\nonumber \\
                             &-& \sum_{j,\nu} \sqrt{\N+j+1}\,a^{(\N+1,s)}_{j\nu} a^{(\N{s})}_{j\nu}~,
\nonumber \\
\sqrt{B}\,D^{(-1)}_{\N-1,\N} &=& \sum_{j,\nu} \sqrt{j+1}\,a^{(\N-1,s)}_{j+1,\nu} a^{(\N{s})}_{j\nu} 
\nonumber \\
                             &-& \sum_{j,\nu} \sqrt{\N+j}\,a^{(\N-1,s)}_{j\nu} a^{(\N{s})}_{j\nu}~.
\label{D_fi}
\eea
The sums include all the terms that can be composed from the 
restricted sets of coefficients $a^{(\N{s})}_{j\nu}$ 
and $a^{(\N\pm{1},s)}_{j\nu}$, while these sets are normalized to 
ensure $\langle \psi | \psi \rangle = 1$ and $\langle \psi' | \psi' \rangle = 1$.

\end{document}